\documentclass[manuscript,screen, acmsmall]{acmart}

\AtBeginDocument{%
  \providecommand\BibTeX{{%
    \normalfont B\kern-0.5em{\scshape i\kern-0.25em b}\kern-0.8em\TeX}}}

\setcopyright{acmlicensed}
\copyrightyear{2024}
\acmYear{2024}
\acmDOI{XXXXXXX.XXXXXXX}

\acmConference[Accepted for publication at CSCW '25]{ACM SIGCHI Conference on Computer-Supported Cooperative Work & Social Computing}{October 18 — 22,
  2025}{Bergen, Norway}
\acmISBN{978-1-4503-XXXX-X/18/06}

\graphicspath{{Graphics/}}
\usepackage{makecell}
\renewcommand{\arraystretch}{1.5}

\usepackage{graphicx}
\usepackage{colortbl}
\usepackage{color}
\usepackage{tabularray}
\usepackage{enumitem}
\usepackage{booktabs}
\usepackage{multirow}
\usepackage{tabularx}
\usepackage{multirow}
\usepackage{booktabs}
\usepackage{xcolor,soul}
\usepackage{subcaption} 
\usepackage{wrapfig}
\graphicspath{{Graphics/}}
\renewcommand{\arraystretch}{1.5}

\definecolor{lightblue}{RGB}{207, 215, 250}

\newcommand{\Npre}{158}
\newcommand{\N}{21}

\definecolor{redf1}{HTML}{FFA6A6}

\usepackage[textsize=scriptsize]{todonotes}

\begin{document}

\title[...]{Measuring, Modeling, and Helping People Account for Privacy Risks in Online Self-Disclosures with AI}

\author{Isadora Krsek}
\email{ikrsek@andrew.cmu.edu}
\orcid{0000-0001-9624-5077}
\affiliation{%
  \institution{Carnegie Mellon University}
  \streetaddress{5000 Forbes Avenue}
  \city{Pittsburgh}
  \state{PA}
  \country{USA}
  \postcode{15213}
}

\author{Anubha Kabra}
\affiliation{%
  \institution{Carnegie Mellon University}
  \streetaddress{5000 Forbes Avenue}
  \city{Pittsburgh}
  \state{PA}
  \country{USA}
  \postcode{15213}
\email{anubhak@andrew.cmu.edu}
}

\author{Yao Dou}
\affiliation{%
  \institution{Georgia Institute of Technology}
  \streetaddress{225 North Avenue}
  \city{Atlanta}
  \state{Georgia}
  \country{USA}
  \postcode{30332}
\email{email@gatech.edu}
}

\author{Tarek Naous}
\affiliation{%
  \institution{Georgia Institute of Technology}
  \streetaddress{225 North Avenue}
  \city{Atlanta}
  \state{Georgia}
  \country{USA}
  \postcode{30332}
\email{tareknaous@gatech.edu}
}

\author{Laura A. Dabbish}
\affiliation{%
  \institution{Carnegie Mellon University}
  \streetaddress{5000 Forbes Avenue}
  \city{Pittsburgh}
  \state{Pennsylvania}
  \country{USA}
  \postcode{15213}
\email{dabbish@andrew.cmu.edu}
}

\author{Alan Ritter}
\affiliation{%
  \institution{Georgia Institute of Technology}
  \streetaddress{225 North Avenue}
  \city{Atlanta}
  \state{Georgia}
  \country{USA}
  \postcode{30332}
\email{ritter.alan@gmail.com}
}

\author{Wei Xu}
\affiliation{%
  \institution{Georgia Institute of Technology}
  \streetaddress{225 North Avenue}
  \city{Atlanta}
  \state{Georgia}
  \country{USA}
  \postcode{30332}
\email{wei.xu@cc.gatech.edu}
}

\author{Sauvik Das}
\affiliation{%
  \institution{Carnegie Mellon University}
  \streetaddress{5000 Forbes Avenue}
  \city{Pittsburgh}
  \state{Pennsylvania}
  \country{USA}
  \postcode{15213}
\email{sauvikda@andrew.cmu.edu}
}

\renewcommand{\shortauthors}{Krsek, et al.}


\begin{abstract}
In pseudonymous online fora like Reddit, the benefits of self-disclosure are often apparent to users (e.g., I can vent about my in-laws to understanding strangers), but the privacy risks are more abstract (e.g., will my partner be able to tell that this is me?). Prior work has sought to develop natural language processing (NLP) tools that help users identify potentially risky self-disclosures in their text, but none have been designed for or evaluated with the users they hope to protect. Absent this assessment, these tools will be limited by the social-technical gap: users need assistive tools that help them make informed decisions, not paternalistic tools that tell them to avoid self-disclosure altogether.
To bridge this gap, we conducted a study with $N=21$ Reddit users; we had them use a state-of-the-art NLP disclosure detection model on two of their authored posts and asked them questions to understand if and how the model helped, where it fell short, and how it could be improved to help them make more informed decisions. 
Despite its imperfections, users responded positively to the model and highlighted its use as a tool that can help them catch mistakes, inform them of risks they were unaware of, and encourage self-reflection. However, our work also shows how, to be useful and usable, AI for supporting privacy decision-making must account for posting context, disclosure norms, and users' lived threat models, and provide explanations that help contextualize detected risks. 
\end{abstract}



\received{16 January 2024}
\received[revised]{16 July 2024}
\received[accepted]{9 December 2024}

\maketitle

\section{Introduction}
Every day, millions of people log onto pseudonymous online fora like Reddit to access a seemingly safe space to vent, seek support, and connect with like-minded others without the additional baggage of their offline identity. Unsurprisingly, prior work has also found that users in these pseudonymous fora are more likely to disclose highly sensitive information \cite{yang2019channel}. While the benefits of these self-disclosures are apparent to users, the potential harms are more abstract and difficult to reason about. This asymmetry can lead to uninformed sharing that can result in regrets \cite{wang2011regretted,sleeper2016everyday}, self-censorship \cite{das2013self,sleeper2013post}, and/or risks of de-anonymization \cite{Afualo_2023, Pulliam-Moore_2021, staab2023beyond}.
As such, a defining dilemma of modern social networking is helping users balance the social benefits they reap through online self-disclosure with the privacy risks associated with those disclosures.

Recent work \cite{canfora2018nlp, zong2019analyzing, morris2022unsupervised, guarino2022automatic, akiti2020semantics} has explored the use of natural language processing (NLP) tools to help users identify potentially risky self-disclosures. However, to date, these tools have not been evaluated with the actual users that they aim to protect. Rather, most of this work has focused on improving model performance in detecting personal disclosures against benchmarks and accuracy metrics that cannot be confidently correlated with what users want \cite{canfora2018nlp, morris2022unsupervised, guarino2022automatic}. Without considering user intentions, needs, and contexts of use, NLP-powered disclosure detection tools run the risk of running headlong into Ackerman's social-technical gap \cite{ackerman2000intellectual}: i.e., the gap between what is possible technically and what is necessary socially. Users often disclose personal information online intentionally: e.g., when seeking emotional or informational support \cite{record2018sought, de2014mental, zhang2018health}, and to build relationships \cite{de2014mental, krasnova2010online}. A disclosure-detection tool that fails to consider \textit{why} users are disclosing personal information, and what threats they're most concerned about in sharing that information, may fall short of helping users make informed decisions.

\begin{figure*}[t]
    \centering
    \includegraphics[width=0.99\linewidth]{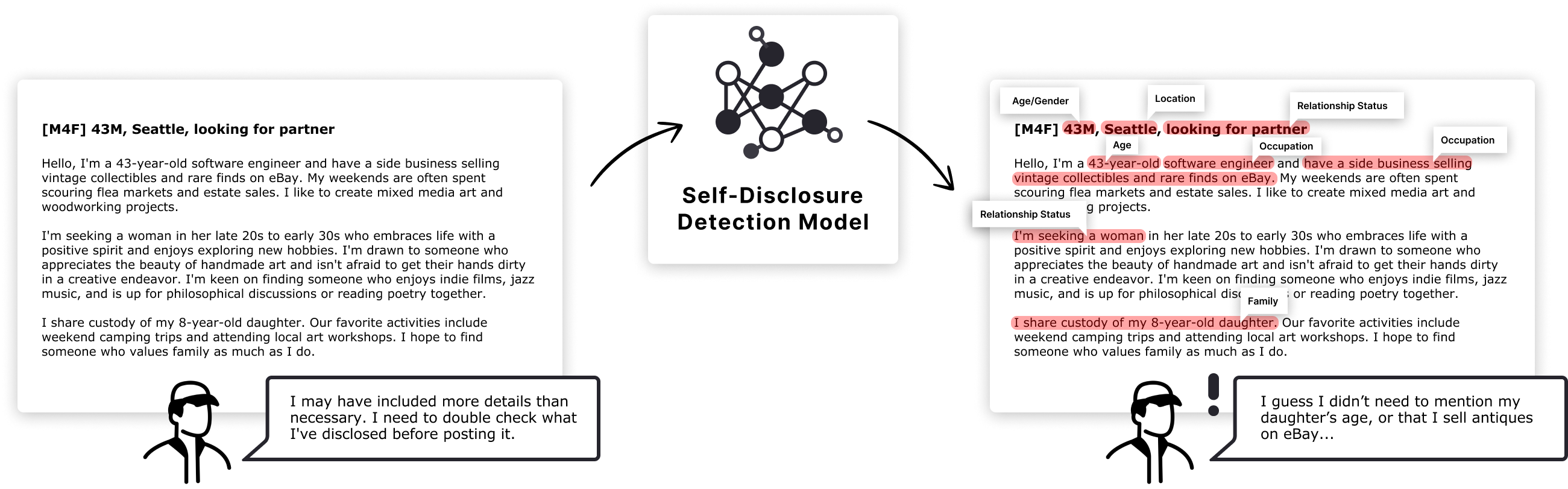}
    \vspace{-8pt}
    \caption{AI-powered, span-level self-disclosure  detection helps users in reviewing their posts to improve privacy.}
    \label{fig:teaser}
\end{figure*}

Driven by this hypothesis, we ask: How can NLP-based disclosure detection tools be designed to help users make informed decisions about online self-disclosures? Taking cues from prior work in the domain of usable security \& privacy, we formulate three sub-questions to better understand how users perceive the outputs of existing models, and how they might be improved:

\begin{itemize}
\item RQ1: Given the tension between the need for self-disclosure and the risks, how can interaction with an NLP model make users aware of potential self-disclosure risks? \cite{acquisti2017nudges, jeung2023correct}
\item RQ2: How should the feedback from the NLP model be framed? What level of textual granularity is helpful in communicating the abstract risks of users' disclosures? \cite{braunstein2011indirect,wang2014field}
\item RQ3: Where do NLP self-disclosure detection models fall short, and how might they be further improved to better align with users' preferences? \cite{taka2023exploring, dalvi2022towards, khritankov2023positive}
\end{itemize}

Guided by these RQs, we conducted an in-depth interview study with {\N} Reddit users. We had our participants use a state-of-the-art NLP-based disclosure detection model from prior work by \citet{dou2023reducing} on two self-authored posts made on and for Reddit. We chose this model because, unlike other similar models that operate at the post or sentence level \cite{valizadeh-etal-2021-identifying,cho-etal-2022-assessing}, Dou et al.'s model operates at the text-span level: i.e., it is capable of identifying specific segments of text in a broader post that constitutes a potentially risky self-disclosure (see Fig.\ref{fig:teaser}).
This higher-level of granularity let us localize user feedback to the specific words in their posts that the model estimated as constituting disclosure risk. Moreover, the model has two versions: a multi-class version (which we refer to as the \textit{categorical disclosure model}) which is capable of detecting 19 different self-disclosure categories (see Table~\ref{tab:disclosure-categories}), and a binary version that flags content as self-disclosures or not (which we refer to as the \textit{binary disclosure model}). To answer our RQs, we asked users questions about how they viewed and accepted the model's outputs for their posts (RQ1), differences in perceptions of model utility between the categorical and binary versions of the model (RQ2), and --- when users rejected model outputs --- why they rejected the outputs and where they felt the model failed more broadly (RQ3).

While the model was imperfect at identifying self-disclosures (with users accepting 58\% of all detected disclosures overall), the majority of participants reacted positively to the model. 82\% of our participants stated that they wanted to use the model outside of the study or would recommend it to others. Participants described the model's utility as encouraging self-reflection, catching users' mistakes, raising risks they were previously unaware of, and in making decisions about content users were uncertain of sharing. However even when agreeing with the model's output, participants generally did not act to alter the disclosure detection spans the model identified (15\% of all detected disclosure spans were altered). Participants rated detected disclosures that they \textit{did not alter} as being especially important to convey meaning. Conversely, participants rated detected disclosures that they \textit{did alter} as being especially risky. In short, our results suggest that model outputs --- even if imperfect --- can help participants make informed decisions: sometimes the risk is warranted, while other times it is not.

We also found that participants preferred the more granular outputs of the categorical disclosure model, as the category labels added clarity to \textit{why} the model detected some disclosures as risky. Category labels served as a kind of ``coarse'' explanation, which participants appreciated. Participants expressed a desire for additional explanation, such as threat severity rankings, worst-case scenarios for how disclosures could be used against users, as well as suggested rephrasings that reduce disclosure risk while retaining the semantic context required for a user's audience. 

Finally, our results indicate many opportunities for future improvements to such models. For example, participants expressed a need for models to understand posting context and disclosure norms (e.g., disclosures about a specific location may be less relevant when posting to a subreddit about that location), to differentiate between factual and non-factual disclosures (e.g., hypotheticals), to account for users' existing privacy management strategies (e.g., perturbing their age, gender, and location when sharing content), and to customize outputs to relevant threat models (e.g., some information may be more exploitable by a stalker than an institution and vice versa). Our findings also surface additional forms of self-disclosure that users found important to detect such as disclosures regarding illegal or abusive activities, substance addiction, and situational details (e.g., dates, times, specific events). 

In summary, this paper makes the following novel research contributions:
\begin{itemize}
\item An examination of how a self-disclosure detection NLP model can raise users' awareness of potential re-identification risks that stem from what they post online in pseudonymous fora.
\item An analysis of how to present model outputs to users to appropriately convey risk in a manner that allows users to make informed decisions about how to weigh the benefits and risks of disclosure when seeking informational and/or emotional support.
\item Design considerations for researchers and practitioners exploring the use of NLP-powered disclosure detection tools in reducing self-disclosure-based privacy risks. 
\end{itemize}

\section{Related Work}
We consider three categories of related work: the current burden of balancing the benefits of self-disclosure with the risks, common approaches from prior work to reduce the risk of online self-disclosure (both non-AI-based and AI-based), and the need for user engagement in the development of AI technology. 

\subsection{The Tension Between the Need for Self-Disclosure and the Risks}
Online pseudonymous communities, like Reddit, offer internet users a safe, seemingly anonymous space to seek opportunities to engage with others free from the baggage of their identity. These interactions often stem from a desire for informational or emotional support, and in seeking support people disclose many personal details publicly \cite{yang2019channel}. Self-disclosure is an important key to establishing a sense of solidarity and community – even disclosing negative experiences has been shown to lead users to feel a sense of support and increased confidence in themselves \cite{semaan2017military}. Online self-disclosure specifically can establish a sense of belonging, and support the development of new relationships \cite{krasnova2010online, cheung2015self} – and is especially valuable for users seeking support on sensitive topics that they might not otherwise be able to receive in their physical communities due to concerns over factors like social stigma, or embarrassment \cite{zhang2018health}. Anonymity when engaging in public self-disclosures can provide users a cover for more intimate and open conversations \cite{bernstein20114chan, stuart2021measure}. 

While self-disclosure fulfills many core social needs and despite users' need to continuously make online self-disclosure decisions based on their intended audiences \cite{yang2019channel}, prior work has shown that people do not always accurately anticipate who might see their content \cite{wang2011regretted}. Balancing informational and emotional support needs with the abstract risks associated with their disclosures (e.g., re-identification, stalking, identity theft, blackmail) is challenging \cite{wang2011regretted, gross2005information}. Unlike face-to-face disclosures, online disclosures leave a digital trace that can be screen-captured and linked to other posts to re-identify and ultimately harm the original posters. To mitigate their risk of re-identification, users employ tactics such as the use of ``throwaway'' or ``burner'' accounts \cite{leavitt2015throwaway, ammari2019self}, but can still end up disclosing enough information about themselves to be re-identified \cite{Pulliam-Moore_2021, Afualo_2023}.

Work on re-identifying users from anonymous datasets shows that few attributes are needed to sufficiently re-identify individuals using incomplete datasets: for example, using 15 demographic attributes, around 99.98\% of Americans can be identified; using just 3, around 83\% can still be identified \cite{rocher2019estimating}. Information as simple as age, gender, and ethnicity can be enough to re-identify an individual when combined with a medical diagnosis \cite{loukides2010disclosure}. Re-identification risks are highly relevant to users of online pseudonymous platforms, like Reddit, with communities dedicated to specific medical diagnoses such as r/Cushings, or r/Marfans wherein users experience many benefits from personal disclosure in seeking support (e.g., earlier detection and treatment of medical issues that would have otherwise gone unaddressed) \cite{joinson2001self, pennebaker2007expressive}.

Relatedly, users' disclosure is guided by their perception of risk and the sensitivity of the information that they are disclosing. It follows that prior research on anonymous public self-disclosure suggests users feel better able to express themselves when posting anonymously, \cite{bernstein20114chan, stuart2021measure}, as anonymity allows users to avoid the face-to-face stigma of embarrassing or uncomfortable topics \cite{donath1999s, de2014mental}. Anonymity affords users the opportunity to be honest, unfiltered, and vulnerable; characteristics they might not be able to exhibit through in-person interactions. Conversely, users' heightened feelings of anonymity can also result in benign disinhibition, wherein users disclose more personal information as they feel more secure and less identifiable \cite{lapidot2015benign}. 

Thus providing users with more awareness of risks to their privacy while maintaining the benefits of disclosures remains a longstanding challenge for security and privacy practitioners. We extend this body of work by exploring the design considerations, utility, and intrusiveness of an end-user-facing NLP-based disclosure detection tool in interactions with users of the online community of Reddit given the impact of pseudonymity on benign disinhibition effects, like increased self-disclosure.

\subsection{Past Approaches to Alleviating the Risks of Self-Disclosure}
\paragraph{Non AI-based Tools}
Researchers and practitioners in security {\&} privacy have explored approaches to helping users navigate the risks of self-disclosures with varying success. One ubiquitous, built-in approach to mitigating users' risks of online self-disclosure is the employment of customize-able privacy preferences \cite{acquisti2017nudges, rathore2017social}. While important, and helpful for users of social networking sites like Facebook, the use of granular privacy settings is not entirely relevant to mitigating the harmful self-disclosure risks pseudonymous online community members face since users are intentionally posting information publicly. Furthermore, despite the protective intention of this, more granularity in privacy settings can actually encourage self-disclosure. In their study, Brandimarte et al. showed that in the context of an online social network, participants who were offered finer-grain privacy controls disclosed more personal information compared to those who were offered weaker controls \cite{brandimarte2013misplaced}. Finer-grain control has also been found to reduce users' privacy concerns in other areas as well such as personalization \cite{taylor2009privacy}. Research has also explored intervention-based strategies for mitigating a user's level of disclosure at the time of posting through the use of framing and presentation. For example, Wang et al. explored the effect of a modified Facebook interface that provided feedback about a post's audience and also allowed users to cancel their post within 10 seconds of making it \cite{wang2014field}. Prior work from Braunstein et al. also showed that changing the wording in a survey question to remind users that they are revealing sensitive information impacts how much they're willing to reveal \cite{braunstein2011indirect}. While these specific techniques are slightly difficult to apply to an online community like Reddit where most posts will be made in public subreddits, they may still be useful for privacy intervention tools. Tools that detect potentially risky self-disclosures for example, may benefit in utility by incorporating additional context for these detections. As such, in this study, we explore the utility of different framing for suggestions by presenting participants with two versions of a disclosure detection model that vary in granularity.

\paragraph{AI-based Tools/existing NLP Approaches to detecting self-disclosure}
With the advancement of AI technology, over the course of the past few years, we've also started to see numerous natural language processing (NLP) based intervention tools emerge as aids for the prevention of privacy leaks from self-disclosure \cite{canfora2018nlp, morris2022unsupervised, guarino2022automatic}. A core benefit offered by NLP-based self-disclosure interventions compared to the aforementioned ones is that they take a more targeted approach to potential privacy leaks by explicitly pinpointing disclosures made by users in their posts. This offers users actionable areas for improvement and the agency to control the dissemination of their sensitive information. Multiple recent works have introduced NLP-based self-disclosure detection models that are targeted towards specific disclosure categories such as medical conditions \cite{valizadeh2023clued, valizadeh2021identifying, zhao2019identifying}, personal opinions \cite{cho2022assessing}, employment history \cite{tonneau2022multilingual}, or stories of sexual harassment \cite{chowdhury2019speak}. Other works developed detection models that consider any type of personal information revealed by users as a single category \cite{reuel2022measuring, blose2020privacy,yang2017self}. A few studies also developed models capable of more fine-grained detection of multiple self-disclosure categories at once \cite{lee2023online, akiti2020semantics}. A major limitation in past work is that it mainly centers around operational changes for enhancing model performance in detecting self-disclosures, such as improving their accuracy or expanding their granularity of categories. However, studying the interaction of users with NLP-based self-disclosure detection models and analyzing user experiences is key in identifying the next steps for the real-world adoption of NLP models in privacy-preserving tools. To our knowledge there has been no prior work on user's reactions to these models, or whether the classifiers used are perceived as being helpful by users. Without the incorporation of user feedback, these NLP detection tools run the risk of ignoring the key benefits of disclosure discussed above. Through seeking out user feedback in the evaluation of AI-assisted self-disclosure detection technology, we hope to surface where these tools may fall short in order to ultimately better align them with the preferences and goals of their intended users. 

\subsection{The Need for User Engagement in the Design of AI Technology}
AI technologies have profound social and economic implications across various domains. When predictive modeling is employed in critical decision-making processes such as credit offers, insurance assessments, hiring, and parole decisions \cite{gries2022modelling, crawford2016ai}, the application of AI can exacerbate social inequality as these technologies have the potential to shape opportunities, leaving certain individuals marginalized due to data or model bias \cite{AI_inequality, plan2016national,Bias_AI_Bloomberg, crawford2016ai}. Additionally, most AI-based technologies are developed from data collected from (biased) human decision-makers, or through training processes which can also introduce bias \cite{taka2023exploring, crawford2016ai}.

It is imperative that the people impacted by AI systems’ deployment be substantively engaged in providing feedback and design direction, and that these suggestions form a feedback loop that can directly influence AI systems’ development and broader policy frameworks which are imperative for ensuring fair and unbiased outcomes \cite{taka2023exploring, khritankov2023positive}. Engaging users in the design of AI technology can help mitigate the negative impacts of AI on society and promote ethical and responsible AI development \cite{crawford2016ai, taka2023exploring}. \citet{park2020scalable} shows the importance of integrating user feedback into the development process of consumer-oriented systems as there is a potential disparity between user expectations and technological functionalities. Recognizing that user preferences and demands can evolve rapidly, the paper advocates for an adaptive approach that places user feedback at the forefront of system design. On a similar note, \cite{dalvi2022towards} talks about the necessity of ongoing updates based on user feedback to prevent errors, particularly in dynamic and evolving scenarios. In contrast to static systems, those that are continually updated and refined based on user interactions demonstrate a heightened ability to adapt to new information, circumstances, and user expectations.

Several works have delved into growing privacy concerns associated with AI-based technologies. The impact of AI on privacy interests is discussed by \citet{kerry2020protecting}, emphasizing how advances in AI amplify concerns regarding the use of personal information \cite{waheed2022empirical, murdoch2021privacy, naithani2024deep, bak2022you}. 

The work of \citet{dou2023reducing} is a first step in assessing user preferences for AI-assisted self-disclosure detection, though it only reports high-level takeaways as to what participants liked and didn't like, and the percentage of users who reported that they would continue to use the model. The goal of Dou et. al.'s paper was to introduce a state-of-the-art disclosure detection model --- the user feedback reporting was primarily to affirm the desirability of their technical contribution to users. In contrast, our goal centers user needs and preferences by aiming to examine: (i) how users make sense of and decide to accept or reject AI-detected self-disclosure risks in their content, (ii) whether and how these AI-detected self-disclosure risks help users make more informed self-disclosure decisions, and (iii) where the model falls short or could be improved in being of greater use to users when they must make these self-disclosure decisions. Our study contributes an exploration of the usefulness, usability, and utility of AI-supported privacy decision-making despite its imperfections.

 Our work ultimately aims to fill this void by examining user perspectives on disclosure and privacy in the area of online communication, contributing knowledge to enhance technical safeguards, and aligning AI systems with user expectations in the rapidly evolving landscape of digital interactions.
\section{Self-Disclosure Detection Model}
Prior works on self-disclosure detection mostly focused on sentence or post-level classification \cite{valizadeh-etal-2021-identifying,cho-etal-2022-assessing} that simply determines if the given text contains any disclosure. However, it doesn't pinpoint the specific spans (i.e., segments of consecutive words) of self-disclosures, providing minimal insights and information for users to review. While there are some models that detect disclosures at the span level, they often have a narrow scope, focusing on a small set of self-disclosure categories \cite{lee2023online} or specialized for news comments \cite{Umar2019DetectionAA}. To bridge these gaps,  \citet{dou2023reducing} recently developed a SOTA self-disclosure detection model that is capable of detecting 17 distinct categories at the word level, which we used as a technology probe in the later user study.

\begin{table*}[t!]
\setlength{\tabcolsep}{2pt}
\resizebox{\linewidth}{!}{%
\renewcommand{\arraystretch}{1.04}
\begin{tabular}{ll}
\toprule
\textbf{Category} & \textbf{Example} \\ \midrule
\textit{\textbf{Demographic Attributes}} \\
\hspace{0.9em}{\sc Age}  & \sethlcolor{redf1}\hl{I recently celebrated my 30th birthday,} and I'm in my final year of grad school. \\
\hspace{0.9em}{\sc Gender} & \sethlcolor{redf1}\hl{I am a guy} and was shamed for doing ballet. \\
\hspace{0.9em}{\sc Age/Gender} & I \sethlcolor{redf1}\hl{(20F)} had a fight with my boyfriend last night. \\
\hspace{0.9em}{\sc Sexual Orientation} & \sethlcolor{redf1}\hl{I identify as a lesbian,} and I'd like to share my thoughts on this matter. \\
\hspace{0.9em}{\sc Husband/BF}  & \sethlcolor{redf1}\hl{My husband} and I plan to travel to Vegas in May. Any recommended hotel?\\
\hspace{0.9em}{\sc Wife/GF} & What's a good birthday gift idea for \sethlcolor{redf1}\hl{my gf}? \\
\hspace{0.9em}{\sc Relationship Status}  & \sethlcolor{redf1}\hl{I've been single} for a while now. \\
\hspace{0.9em}{\sc Race/Nationality}  & \sethlcolor{redf1}\hl{I was a Swedish tourist} in France, once. \\
\hspace{0.9em}{\sc Location} & \sethlcolor{redf1}\hl{I have just moved to United States,} and I'm finding the medical costs quite high. \\
\hspace{0.9em}{\sc Pet} & \sethlcolor{redf1}\hl{I recently got a Labrador Retriever puppy,} and it's been very fun \\
\hspace{0.9em}{\sc Appearance} & \sethlcolor{redf1}\hl{I have bright red hair,} which always draws a lot of attention. \\
\hspace{0.9em}{\sc Name} & Hello guys, \sethlcolor{redf1}\hl{my name is xxx.} I'm looking for a Chinese name for myself. \\
\hspace{0.9em}{\sc Contact} & \sethlcolor{redf1}\hl{My ins is xxx.} \\
\addlinespace[2.4pt]
\textit{\textbf{Personal Experiences}} \\
\hspace{0.9em}{\sc Family} & \sethlcolor{redf1}\hl{My brother (23M)} fought my husband (30M) at a family dinner. \\
\hspace{0.9em}{\sc Health}  & \sethlcolor{redf1}\hl{I've diagnosed with a prostate cancer}, but good news is it's early stage. \\
\hspace{0.9em}{\sc Mental Health} & \sethlcolor{redf1}\hl{I have bipolar, anxiety, PTSD and ADHD,} which are parts of me. \\
\hspace{0.9em}{\sc Occupation} & For context, \sethlcolor{redf1}\hl{I'm a L5 software engineer at Google.} \\
\hspace{0.9em}{\sc Education} & \sethlcolor{redf1}\hl{I just completed my PhD} today, and I'm feeling incredibly happy about it! \\
\hspace{0.9em}{\sc Finance} & This year, \sethlcolor{redf1}\hl{I made 100K from Bitcoin,} which I used to purchase a new car. \\ \bottomrule
\end{tabular}
}
\caption{Examples for each of the 19 self-disclosure categories, grouped into \textit{Demographic Attributes} and \textit{Personal Experiences}. The categories presented in the above table are those that are detectable by Dou et. al's model \cite{dou2023reducing}, while the examples used have been drafted by the authors of this paper.}
\label{tab:disclosure-categories}
\vspace{-15pt}
\end{table*}

The model demonstrates high performance with 75\% F$_1$ in the categorical disclosure setting (classify each word into one of 17 categories or as non-disclosure) and 82\% F$_1$ in the binary disclosure setting (classify into disclosure or non-disclosure). We experimented with both versions to explore how different levels of label detail affect user experience --- per RQ2. Based on RoBERTa-Large \cite{liu2019roberta} transformer encoder, the model is fine-tuned on a Reddit corpus containing 4.8K human-annotated disclosure spans. We chose this model over prompting-based LLMs such as GPT-4 \cite{OpenAI2023GPT4TR} due to its superior efficiency and cost-effectiveness, as well as its similar performance: as \citet{ashok2023promptner} shows fine-tuned small models generally outperform zero-shot prompting of larger models on span annotation tasks.

\begin{figure*}[htp]
\vspace{-0.8 \baselineskip}
 \centering
    \includegraphics[width=8cm]{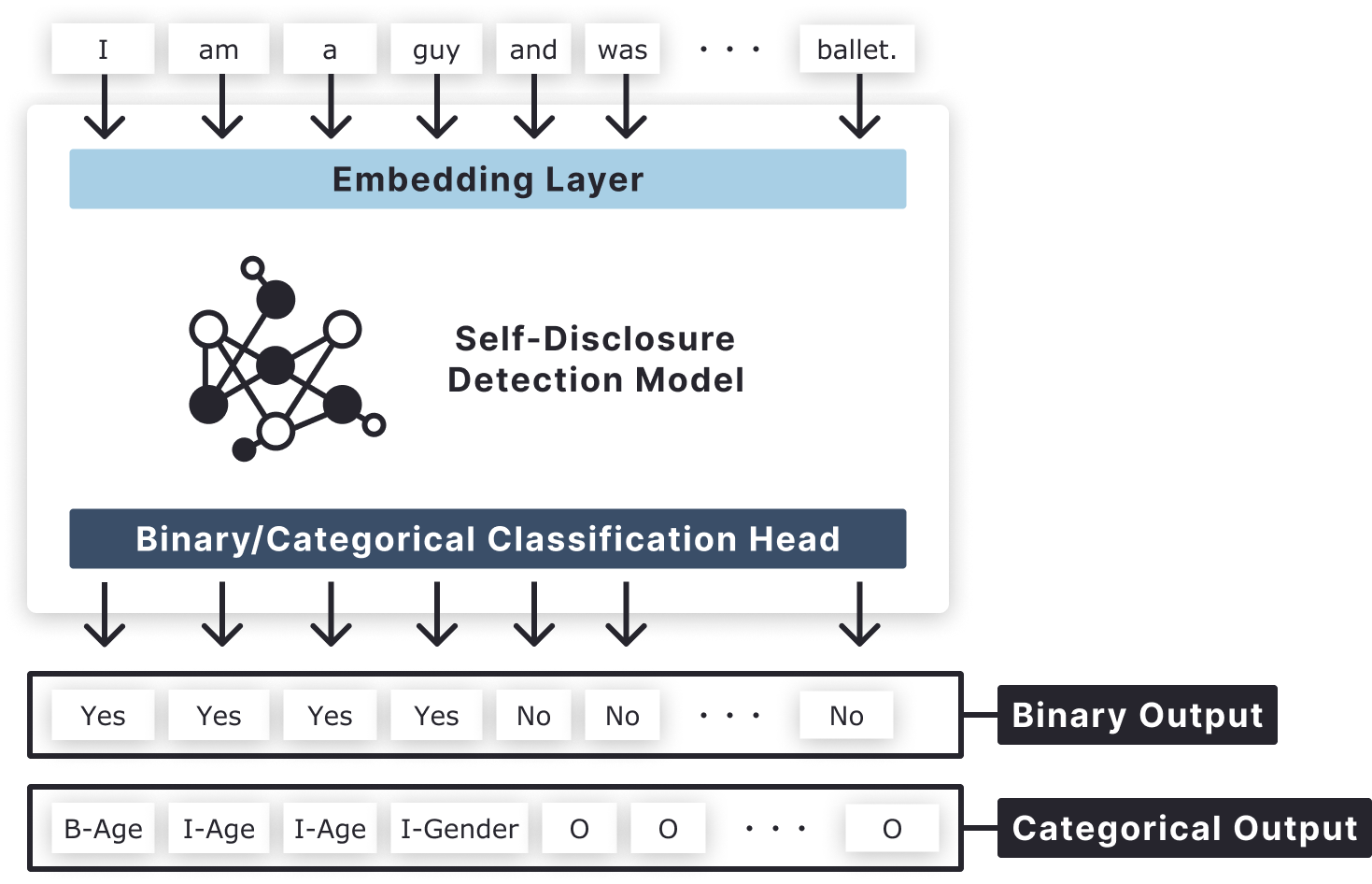}
   \vspace{-8pt}
    \caption{The model classifies each word to a label.}
   \label{fig:model_diagram}
  \vspace{-1.1\baselineskip}
\end{figure*}

During inference, instead of feeding the entire post, including both title and body, into the model, we first used Ersatz \cite{wicks-post-2021-unified}, a neural network-based sentence splitter, to divide posts into individual sentences. Each sentence is then fed into the model, resulting in better performance over processing the whole post at once. To be more specific, for a sentence of $n$ words $x_1, ..., x_n$, the model assigns a corresponding label $y_i$ to each word $x_i$.
In the categorical disclosure setting, we used the IOB2 label format \cite{tjong-kim-sang-veenstra-1999-representing}, which tags the starting word of a disclosure span as \textit{B-[Class]} and subsequent words as \textit{I-[Class]} (e.g., B-Age, I-Age), while non-disclosure words as \textit{O}. For the binary disclosure setting, the labels are simply \textit{yes} or \textit{no}. As words are tokenized, we use the label of the first token of each word as a word label. Figure \ref{fig:model_diagram} illustrates the model diagram.

In addition to the 17 main categories detected by the model --- ranging from age to mental health, frequently appearing on Reddit --- we also addressed the identification of less common but sensitive details like names and contact information. To identify names, 
we used LUKE \cite{yamada2020luke}, the state-of-the-art named-entity recognition model. For the detection of personally identifiable information, such as phone numbers and emails, we integrated Microsoft Presidio\footnote{\url{https://microsoft.github.io/presidio/}}.
Recognizing that social media handles are a commonly shared type of contact information on Reddit, we extended Presidio with custom regular expressions to detect usernames from platforms such as TikTok, Instagram, and Twitter.
Given our emphasis on self-disclosures, we applied the sentence classifier from \citet{dou2023reducing} as a preliminary self-disclosure filter before using these external tools:
The classifier first identified whether a sentence contained self-disclosure elements; sentences flagged by this classifier were then subsequently processed through these external models and tools.

Table \ref{tab:disclosure-categories} presents examples for all 19 categories, which are broadly grouped into: \textit{demographic attributes} and \textit{personal experiences}.
Demographic attributes include static personal characteristics, typically expressed in a brief manner, such as age, gender, and race.
In contrast, personal experiences relate to activities and events that might identify an individual --- e.g., health-related and education-related activities or questions.

Of note, our goal in this work is \textit{not} to introduce new state-of-the-art self-disclosure detection models. Rather, we use these models as a technology probe \cite{hutchinson2003technology} in a study to understand the opportunities and challenges of using AI to help users make informed decisions when sharing personally identifiable information online.
\section{Methodology}
To explore how AI-assisted identification of self-disclosure spans might impact users' awareness of self-disclosure risks and their writing behaviors, we conducted a 75-minute interview study with 21 Reddit users. We asked participants to share two of their self-authored Reddit posts with researchers which were subsequently run through our NLP model. We then asked participants questions, using the NLP model outputs as a guide, to determine how the model affected users' awareness of potential self-disclosure risks (RQ1), how feedback from the model should be framed to maximize utility to participants (RQ2), and where these tools might be altered to better align with users' preferences (RQ3). 

\subsection{Recruitment}
 We recruited a total of {\N} Reddit users through Prolific, an on-demand participant recruitment platform. Participants were screened to ensure that they: had a Reddit account at the time of the study, made at least three posts on Reddit, resided in the U.S., and were 18 years of age or older. Eligible participants were subsequently asked to provide links to two text-based Reddit posts they had made; one with which they had privacy concerns and another with which they did not. From about {\Npre} who took the pre-study survey, we invited 33 Reddit users to participate in the interview based on the Reddit posts they shared with which they had privacy concerns. Participants were only invited to participate after researchers ensured their shared posts were majority text-based, and contained personal disclosures, and after ensuring that the model we used was able to detect potentially identifying disclosures in the post. More specifically, participants were filtered out of the study if the posts that they shared were inaccessible, removed, contained no text (e.g. only an image), or if they reported not having made a post that meets those criteria (e.g., if they answered with ``N/A'' or ``I have not done this'' when asked to share a Reddit post they drafted over which they had privacy concerns). While we did reach out to participants who shared invalid links, none of them responded to our request for an alternative post. From the total {\Npre} users who took the pre-study survey, we invited 33 Reddit users to participate in the interview based on the Reddit posts they shared with which they had privacy concerns. Of those 33 participants, {\N} followed up on our request to schedule interviews. Given the average sample size for studies within the CHI community is 12 participants \cite{caine2016local}, and the rich body of work demonstrating that anywhere between 6-12 participants is enough for high saturation \cite{morgan2002risk, guest2006many, namey2016evaluating, francis2010adequate, hagaman2017many, guest2020simple}, we did not run any additional recruitment attempts.

The collected demographic data is displayed in the Appendix (see Table A in section A). The sample was slightly skewed female, with 12 participants identifying as female. 16 of the 21 participants were below the age of 50, and 15 participants held a bachelor's degree or higher certificate of education. Data collection occurred in the summer of 2023, and participants received {\$}17 compensation for their participation in the interview and pre-study survey. Our study design was approved by an institutional review board. 

\subsection{Pre-study Survey}
Participants were first directed to an online survey that included screening questions to ensure eligibility. Those who didn't meet the requirements were automatically removed from the survey at this stage. Afterward, participants were presented with a consent form in the survey. Those who agreed to participate provided their preferred email address for future communication and to schedule a Zoom-based interview. Participants were then requested to share links to two Reddit posts they had made; one with which they had privacy concerns and another with which they did not. As mentioned prior, participants were filtered out of the study if the posts that they shared were inaccessible, removed, contained no text (e.g., had only an image), or if they reported not having made a post that met our criteria.

In the final section of the pre-study survey, participants answered questions about their Reddit usage and behaviors, their perception of self-anonymity on Reddit, as well as their tendency to disclose information publicly using validated scales adapted from prior work on perceived anonymity in online social support communities \cite{yun2006creation}. Additionally, participants were asked about their demographics and their general attitudes towards security and privacy using the SA-13 scale \cite{faklaris2022they}. We collected these data because prior research has found that differences in end-user security attitudes, demographic characteristics, and perceived audiences when sharing content online can shape users' online S{\&}P decision-making \cite{chen2012students}.

\begin{figure*}[htp]
    \centering
    \includegraphics[width=15cm]{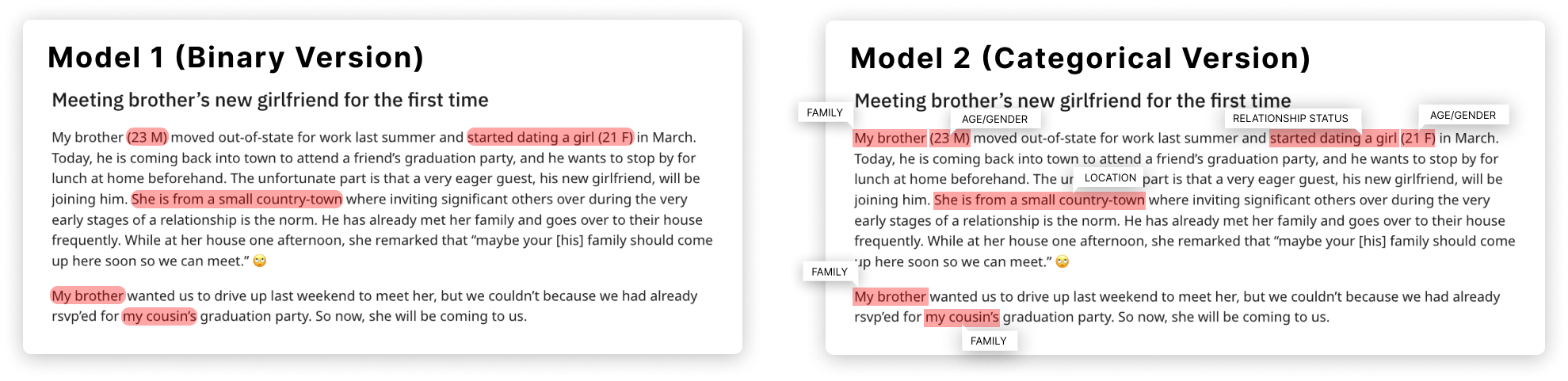}
    \caption{Image depicting how the disclosure detection spans of both versions of the model were presented to participants. On the left is the binary disclosure version, and on the right is the categorical disclosure version.}
        \Description{Image of how both the disclosure detection spans were displayed to users for both versions of the model. On the left, the binary disclosure model is displayed: for this version suspected self-disclosures were highlighted in red. On the right, there is an image of the same Reddit post with the categorical disclosure model detections: here suspected self-disclosures are also highlighted in red, and additionally labeled with a category expressing the type of self-disclosure detected by the Categorical disclosure model.}
    \label{fig: Presentation of Self-Disclosure Spans By Model Type}
\end{figure*}

\subsection{Interview}
\paragraph{User Assessment of Model Output on Their Posts}
We conducted 75-minute semi-structured interviews to gather feedback on the model's outputs and understand participants' views on online anonymity risks and how those risks are influenced by subreddit community norms. The interviews began with general questions about participants' Reddit usage and the disclosure norms they followed for the subreddits in which they were active. We also asked about their posting behaviors and differences in disclosure norms across subreddits in order to account for posting practices that might interfere with the outputs of the model (RQ1, RQ3). Prior to showing participants the model outputs on their posts, we discussed the posts participants shared with us in the pre-study survey. We started with the post they reported feeling less comfortable sharing with people they know in the physical world, probing them on what personal disclosures they thought they included in the post, as well as the threats they perceived to their anonymity, who they were concerned might identify them, and why. We then asked participants to either (a) share a post from a burner account they had privacy concerns with, or (b) draft a post that they'd previously hesitated to share on Reddit for reasons related to privacy or sensitivity. After doing this, we then probed participants on all the same questions as the first post. As opposed to the first post, asking them to share/draft an additional post allowed us to assess the model's utility in providing feedback on posts that users hesitated to share from their main accounts. 

In the next stage of the interview, we analyzed participants' posts using both the binary and categorical disclosure versions of the disclosure detection model (RQ2). Participants first saw the output of the binary disclosure model on both of their posts and assessed those. We then showed participants the output of the categorical disclosure model on their posts for assessment. When evaluating the model's outputs on self-authored posts, we also asked participants to assess each of the disclosure spans detected by the model; we asked whether they agreed or disagreed that the detected disclosure span contained self-disclosure and why, as well as whether and how they would want to change their post as a result. Participants then rated each detected disclosure span on four aspects using a 5-point Likert scale (from ``not at all...'' (1) to ``very...''(5)): 1) the ``helpfulness'' of the disclosure span in surfacing a privacy risk, 2) the ``importance'' of disclosing the information in the disclosure span for communicating the context of the post, 3) their perceived ``sensitivity'' of the information in the disclosure span, and 4) the ``riskiness'' of disclosing that information in their online post. Because these scales were delivered verbally in during the interview, we also probed participants on the reasoning behind their rankings. We present our analysis of their reasoning in the results section alongside rankings. Finally, we asked participants whether they would feel comfortable posting the modified second post to their main Reddit account based on the model outputs, and any changes they would make as a result (RQ1, RQ2, RQ3). 

In the next stage of the interview, in order to get a baseline user assessment for each of the category labels for disclosure spans produced by the categorical disclosure model, we asked participants to assess the disclosure detection outputs of the model on a selection of public posts sourced from Reddit (RQ2). We selected six such posts where the model performed reasonably well; each participant was randomly shown one of the 6. Upon being presented with the disclosure detections from the model on the public post, participants then answered all the same questions they were asked in their assessment of the model output on their self-authored posts.

Lastly, after completing all assessments of the model outputs we asked participants to share their overall opinions on the model (both positive and negative) (RQ1), what changes could be made to improve its utility (RQ2), and whether or not they would use it themselves and why (RQ1, RQ3).

\begin{figure*}[htp]
    \centering
    \caption*{INTERVIEW METHODOLGY FLOWCHART}
    \includegraphics[width=14cm]{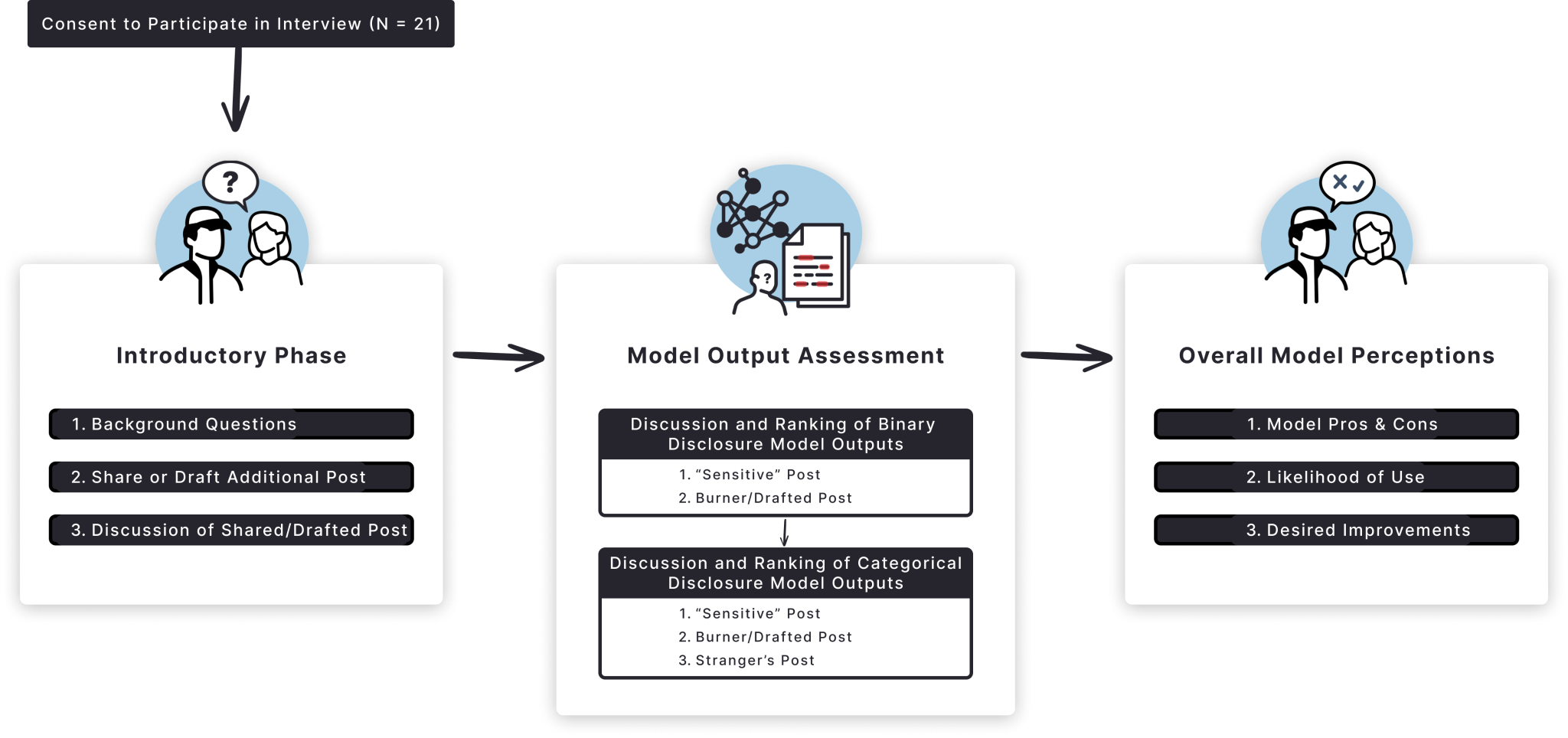}
    \caption{Flowchart depicting the interview methodology flow. Participants first answered general questions about self-disclosure on Reddit and were asked to share a post from a burner account or draft a post that they considered making but hesitated to due to privacy concerns. They then moved onto the model output assessment stage where they provided their thoughts on the model's output across the two posts they shared and ranked them on helpfulness, importance, riskiness, and sensitivity – first assessing the outputs of the binary disclosure model, and then the categorical disclosure model. Finally, the participants then moved to the exit stage of the interview, being probed on their overall thoughts on the model and concerns, as well as desired areas for improvement.}
        \Description{Flowchart depicting the interview methodology flow. It flows from The interview introduction stage, which included 1. general questions about users' Reddit use behaviors and disclosure on Reddit, 2. Users were given a set amount of time to either share a post that they made to a burner account or were asked to draft a post they had considered making but didn't over privacy concerns, and 3. Discussion of the posts that users shared (both the ``sensitive'' and the burner/drafted post). Next users engaged in the model output assessment stage wherein users provided their thoughts on the model's output across the two posts they shared and ranked them on helpfulness, importance, riskiness, and sensitivity – first assessing the outputs of the binary disclosure model, and then the categorical disclosure model. Finally, the participants then moved to the exit stage of the interview, being probed on their overall thoughts on the model and concerns, as well as desired areas for improvement.}
    \label{fig: Presentation of Self-Disclosure Spans By Model Type}
\end{figure*}

\subsection{Data Analysis}
We used an inductive thematic analysis approach to coding our interviews with participants \cite{proudfoot2023inductive}, leveraging affinity diagramming to uncover themes within the work. This analysis was conducted by two researchers who actively collaborated in reviewing 5 transcripts to develop a codebook with 40 codes, containing a mixture of general reactions to the model, disclosure detection preferences, and their reasons for each. The same two researchers then used this codebook to identify emergent patterns and themes across the full set of interview transcripts. The researchers also collaborated in affinity diagramming the codes in order to reveal sub-themes in the work. For a subset of the study sessions (23\% of the data) which were coded independently by two researchers, we achieved a Cohen’s Kappa of 0.98 --- an inter-rater reliability value generally considered to represent high agreement \cite{landis1977measurement, viera2005understanding, fleiss2013statistical}.

In order to analyze how participants' reactions to the disclosure detections surfaced by each model differed, we compared:
The rate at which participants agreed that the detected disclosure span contained a sensitive self-disclosure; the reasons participants provided for disagreeing with or rejecting a disclosure span; whether participants altered what they wrote based on the disclosure detection by the model; and, the rating of each detected disclosure span in terms of its helpfulness (of the disclosure detection span in surfacing a privacy risk), importance (of including in the post for context), sensitivity of the information disclosed, and the riskiness of sharing that information.

To explore whether participants' rankings (of helpfulness, importance, sensitivity, and riskiness) differed significantly across altered and non-altered disclosure spans that were accepted by participants as containing self-disclosure, we employed both a two-sample t-test and Mann-Whitney test to ensure the results agreed to account for any power differences. The p-values reported are from the two-sample t-test. We ran all statistical comparisons in R.
\section{Results} 
To answer RQ1, our findings indicate that users found value in using our model to detect self-disclosure risks, with the majority (17/21) expressing a desire to use it outside of the study or recommend it to others. In particular, users appreciated the self-reflection the model encouraged, finding value in its ability to: (i) catch risky disclosures they were previously unaware of or missed in the process of writing, (ii) broaden their conception of risky disclosure, and (iii) help them come to a decision on information they were unsure about disclosing. However, we also found that users required support in navigating how to rephrase and de-risk the content that they wrote in a manner that preserved the semantic meaning of their original post.
To answer RQ2, our results suggest that the granularity of information offered alongside text that the model detected as risky may have had an impact on users' acceptance of disclosure detection spans (the text that the models determined may contain self-disclosure risks). Users described the categorical disclosure version of the model's category labels as helping them understand \textit{why} a disclosure span was detected as a privacy risk.
Finally, to answer RQ3, our results indicate that participants found the model least useful when its outputs failed to account for their own risk mitigation strategies (e.g., hypothetical situations, intentional lies users made to preserve their anonymity), conflicted with the disclosure norms and requirements of their posting context, and surfaced risks that were misaligned with the threat models they most cared about. We also highlight categories of risky disclosure that the model failed to capture, but which participants felt were important to detect.

\paragraph{Descriptive Statistics on Users' Reactions to the  Self-Disclosure Detection Spans Surfaced by the Model}
Among the 851 total self-disclosure spans that the model detected across all our participants, participants accepted 58\% (495 total) as containing risky self-disclosures, and rejected the other 41.8\% (356 total) (see Table 3). In response to seeing the detected disclosures, about 15\% (127/851) of all disclosure spans were ultimately altered by participants, while 3.5\% (30/851) left participants undecided on what to do; the remaining 81.5\% (694/851) were not altered by participants. 

When only looking at the disclosure spans that were accepted by users (495), the alteration rate of disclosure spans was only slightly higher (114/495 = 23\%), with the majority still deciding not to alter the disclosure span (357/495 = 72\%); participants were unsure whether or how they would alter the remaining accepted disclosure spans (23/495 = 5\%).

To explore whether participants' decisions to alter the content of detected disclosure spans were influenced by mistakes made by the model, we further examined the 72\% of non-altered disclosure spans that participants accepted as containing self-disclosures (357). For these detected disclosures, we calculated the rates at which participants reported different mistakes by the models. We found that very few of the non-altered disclosure spans were reported to have contained mistakes made by the model (10/357 = 3.2\%). Only about 7 (7/357 = 2.2\%) were identified by participants as misaligned with users' preferences for what to disclose, while only 3 (3/357 = 1\%) had mistakes pertaining to post-context (e.g., the model flagging disclosures that users were required to share per subreddit rules). The majority of non-altered but accepted disclosure spans contained no explicit mistakes identified by users (305/357 = 85\%).

Rather, participants' decisions to alter text seemed related to their perceptions of the disclosure span's helpfulness, importance, and perceived riskiness. While there is debate over how to analyze Likert scale data, past work in statistics has shown that two-sample t-tests and Mann-Whitney U tests produce nearly equal false-positive rates when performed on Likert scale data \cite{de2010five}, and are close in significance level target. Given this reported equivalence we conducted a Welch's t-test in order to examine whether the medians of the altered (114) and non-altered (357) disclosure spans that participants agreed with (495) differed significantly across scores of helpfulness, importance, sensitivity, and riskiness (as it accounts for unequal variances, and is considered to be more conservative than student t-tests and Mann-Whitney U \cite{ergin2023comparison}); we found significant differences across all four scales (see Table 2). \footnote{As a robustness check, we also performed a Mann-Whitney U test, another non-parametric test of statistical difference often used on Likert data and obtained the same results. Conducting both tests in assessing Likerts is understood to be helpful in cementing confidence in purported significance levels \cite{de2010five}. See Appendix Section A.5 for a full analysis.}

From our tests, we found that the content of non-altered disclosure spans was perceived as more ``important'' to include for the context of the participant's post than that of altered spans (t(155)=7.244, p < 0.0001, Cohen's d = -0.4545076). In contrast, the content of altered spans was considered more ``sensitive'' (Welch's: (t(182)=8.928, p < 0.0001, Cohen's d = 0.5217734))) and ``risky'' by participants (t(168)=8.7505, p < 0.0001, Cohen's d = 0.5287747) than that of non-altered spans. Unsurprisingly, therefore, the model highlights for altered spans were perceived as being more ``helpful'' (t(201)=6.9965, p < 0.0001, Cohen's d = 0.2475993) than the highlights for unaltered spans. We use the subsection on RQ3 below to extend our discussion of the reasons participants gave for rejecting disclosure spans output by the model. 


\begin{table}[htp]
\centering
\caption*{SUMMARY STATISTICS FOR ACCEPTED SELF-DISCLOSURE SPANS OUTPUT BY THE MODELS}
\resizebox{\linewidth}{!}{%

\begin{tabular}{lcc|cc|ll} 
\hline\hline
\multicolumn{1}{c}{} & \multicolumn{2}{c|}{\begin{tabular}[c]{@{}c@{}}Accepted Altered Disclosure Spans\\ (114)\end{tabular}} & \multicolumn{2}{c|}{\begin{tabular}[c]{@{}c@{}}Accepted Non-Altered Disclosure Spans \\ (357)\end{tabular}} & \multicolumn{2}{c}{Welch's T-Test Results}  \\ 

\cline{2-7}
                     & M    & SD                                                                                              & M    & SD                                                                                                   & p-value      & Cohen's \textit{d }          \\ 
\hline
Helpfulness          & \texttt{4.14} & \texttt{1.24}                                                                                            & \texttt{3.15} & \texttt{1.44}                                                                                                 & \texttt{1.52702e-10***} & \texttt{0.24}                    \\
Importance           & \texttt{2.58} & \texttt{1.53}                                                                                            & \texttt{3.74} & \texttt{1.25}                                                                                                 & \texttt{7.728284e-11***} & \texttt{0.45}                    \\
Sensitivity          & \texttt{3.95} & \texttt{1.36}                                                                                            & \texttt{2.62} & \texttt{1.39}                                                                                                 & \texttt{1.884628e-15***} & \texttt{0.52 }                  \\
Riskiness            & \texttt{3.61} & \texttt{1.47}                                                                                            & \texttt{2.22} & \texttt{1.35}                                                                                                 & \texttt{8.741327e-15***} & \texttt{0.52}                    \\
\hline
\end{tabular}
}
\caption{Welch T-test results for disclosure spans accepted by participants to explain \emph{nuances in when participants altered and did not alter their posts} in response to seeing the model outputs. The outcomes of the test show significant differences in the
median scores of disclosure spans between those accepted and altered compared to those accepted but not altered by participants across helpfulness, importance, sensitivity, and riskiness. P-values for both tests have been adjusted for multiple-hypothesis testing. 
\\Helpfulness, Importance, Sensitivity, and Riskiness all range from 1 (“Not at all...”) to 5 (“Very...”) \\ *\emph{p}<0.05, **\emph{p}<0.01, ***\emph{p}<0.001
\\M = Mean, SD = Standard Deviation
}
\end{table}

\subsection{RQ1: How Interaction with an NLP Model Can Benefit Users' Awareness of Risky Self-Disclosures and Mitigation Behaviors}
While the model was clearly imperfect, when reflecting on their overall opinion of the tool, participants' reactions to the model were largely positive:
The vast majority of participants (17/21) either expressed that they would want to use it personally when making online posts (14/21), or that they would want to recommend its use to others they know 2/21. That latter group felt like they already had appropriate strategies for mitigating self-disclosure risks in their posts without the model, but suggested it might be especially useful for users with less experience in mitigating disclosure risks online. P19, for example, recounted: ``I don't know that I necessarily need it. I do think it'd be a good idea for like tweens and teens, like people who are new to the internet. Just kind of give them an idea of stuff that they really shouldn't be sharing.'' 

Users perceived several benefits from their interaction with the model, with most describing its value as a tool for reflection (12/21), even for circumstances where the model's outputs made little sense in isolation. Participant's responses revealed that a core value of the model lay in its potential for catching mistakes and double-checking one's writing for potential privacy risks:  ``What it grabs prompts you to think more thoroughly about whether you actually want to include that information, and whether that information could be used to identify you. Specifically, in terms of r/AmItheAsshole\footnote{r/AmItheAsshole is a subreddit that Reddit users post to in order to get outside perspective on conflicts they've experienced, and whether the way that the author handled that conflict was justified.} posts --- the number of times I've seen people edit their post with 'so and so found it'...'' (P8)

Other participants mentioned the utility of the model as a ``tie-breaker'' when deciding whether or not to disclose specific information, as evidenced by the following reflection from P20: ``it would help me make the decision of `do I want to take this out, or change it, or is it fine'...''.

Finally, other participants mentioned the model's potential to challenge and subsequently expand users' conception of what constitutes a risky self-disclosure: ``It catches very detailed things that people may write that they may not even think about as being risky, like things about location that people reveal not even realizing that they're revealing it sometimes. Sometimes we're not even realizing it could be risky.'' (P5)

\subsection{RQ2: The Impact of Classification Granularity}
To assess how the granularity of classification labels might impact the utility of the disclosure detection spans from the model to users, we also assessed the differences in users' reactions to a binary disclosure model (where text spans are classified as risky self-disclosures or not) versus as a categorical disclosure model (where these spans were also labeled with a category describing the kind of sensitive disclosure made). 

Overall, we found that higher-granularity classifications resulted in greater user acceptance of model outputs: outputs from the binary disclosure model were rejected (54\%) more often than they were accepted (46\%), while outputs from the categorical disclosure model were accepted (65\%) more often than they were rejected (35\%).

\begin{figure*}[htp]
    \centering
    \caption*{BARPLOT OF ACCEPTED VS. REJECTED DISCLOSURE DETECTION SPANS, BY MODEL TYPE}
    \includegraphics[width=13.5cm]{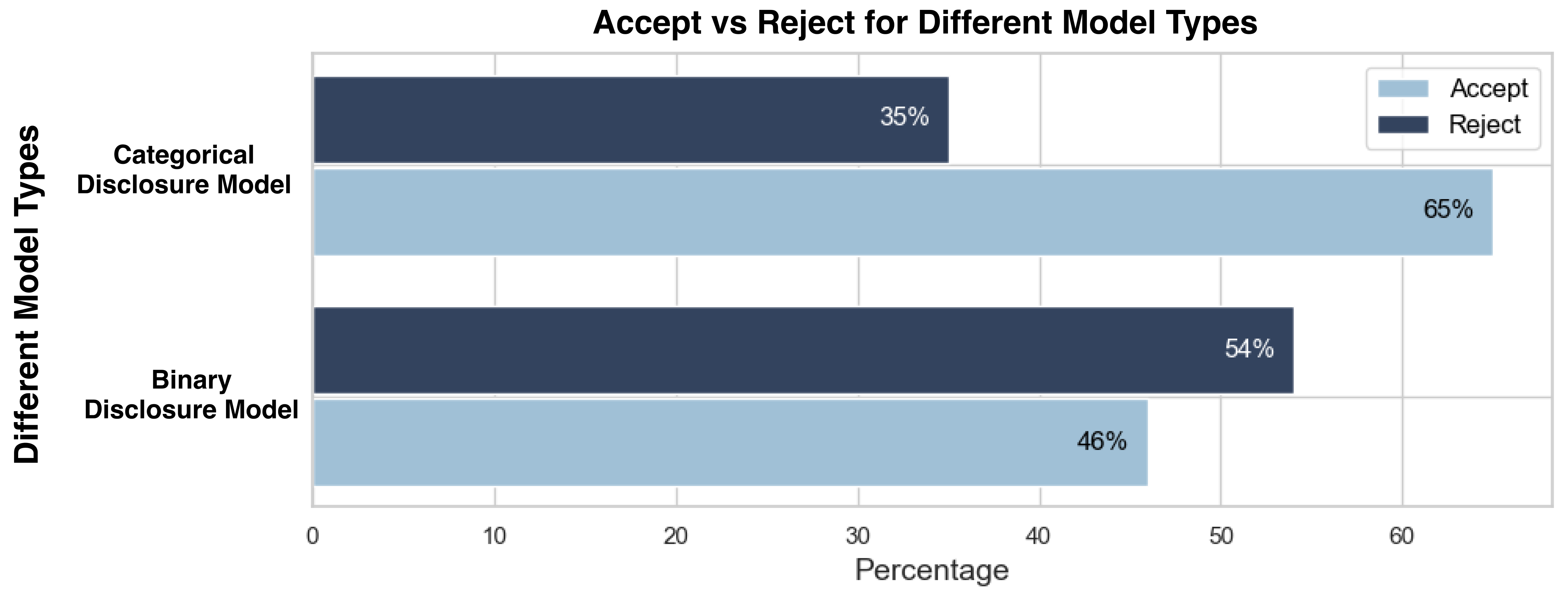}
    \caption{A horizontal bar graph depicting the percentage of disclosure detection spans that participants accepted vs. rejected as containing self-disclosure risks, by model type. The percentage of disclosure spans that were accepted output from the binary disclosure version of the model (46\%) was lower than those accepted (54\%). Compared to the binary disclosure version, we see that for the categorical disclosure version of the model disclosure spans were accepted more often, with the majority of participants (65\%) agreeing they contained self-disclosure risks.}
        \Description{A bar graph depicting the difference in self-disclosure detection span acceptance and rejections by model type on the Y-axis, against number pauses from 0 to 70 in increments of 10 on the X-axis. 4 lines are shown. Disclosure detection spans from the categorical disclosure model have the highest rate of acceptance (65\%). Disclosure detection spans from the binary disclosure model have the lowest rate of acceptance, with the majority of binary disclosure spans being rejected (56\%).}
    \label{fig: Presentation of Self-Disclosure Spans By Model Type}
\end{figure*} 

However, greater acceptance of disclosure spans did not necessarily result in markedly different disclosure behaviors. We define the ``alteration rate'' of a model as the number of detected disclosures that users decided to alter, divided by the total number of disclosures detected by that model. 
The alteration rates for both the categorical disclosure (91/557 = 16.4\%) and binary disclosure (36/289 = 12.5\%) models were both similar ($\chi^2(1, N=846)=1.9, p=0.16$) and low overall. Note, however, that the detected disclosures identified by the binary disclosure model and the categorical disclosure model were not always the same, so we cannot make direct comparisons at the text-span level between the models.

Higher-granularity labels in the categorical disclosure model helped participants (10/21) recognize the disclosure risks of a detected text span. The presence of these categories helped surface \textit{why} a span of text was detected by the model, which was instructive in situations where participants may have not been aware of a specific type of risk. As P21 stated, ``I would have never thought that would have been...identifiable information...but gender is...I mean, obviously with outliers, but there's the binary of gender in terms of sex of birth which you've now basically taken off the table.''

The average ``helpfulness'' and ``riskiness'' scores that participants assigned to detected spans also differed slightly between the two models. Recall that participants were asked to rate the perceived helpfulness and riskiness of each detected span on a 5-point Likert scale (ranging from ``not at all...'' to ``Very...''). The detected spans of the categorical disclosure model were, on average, perceived as ``slightly helpful''/``helpful'' and ``slightly risky''. In contrast, the detected spans of the binary disclosure model were perceived as ``slightly helpful'', but ``not at all'' risky. P19 discussed how the higher granularity of the categorical disclosure model outputs may have contributed to the improved perceptions of helpfulness and agreement of risk: ``I think that [the categories] would be very helpful...it might not be obvious why something is considered a disclosure. Like, why would they highlight this? What? What's revealing about this, and maybe having the category would be like that? Oh, yeah, I guess I see the point.'' In this way, the categorical disclosure labels can be seen as a kind of ``coarse'' explanation for model outputs.

Relatedly, the majority of users expressed a desire for more granular explanations for model outputs (16/21). Participants wanted explanations that spoke to why a detected disclosure was considered a risk, how sensitive or risky was the detected disclosure, and worst-case scenarios or examples of how the information in the detected disclosure could be used to harm them. Participants also expressed a desire for feedback on how to modify the detected disclosure to reduce risk while retaining enough context for their desired audience to understand their posts. 

Participants discussed wanting the model to take their history of disclosures across posts into consideration when detecting and explaining potential risks (6/21). As P14 stated: ``It might be also important to point out like `combined with other information in your posts plus other information that may be in your profile on previous posts, this sort of thing may be identifiable!''' (P14). P20 also expressed a desire to contextualize model outputs based on how other users have historically responded to similar outputs: ``It could tell me like statistics about other users and like tell me what other people decided based on the information that it like gathered for them. Maybe that would help me make the decision of do I want to take this out or change it or it's fine'' (P20). 

\begin{table}[htp]
\caption*{SUMMARY OF MODEL DISCLOSURE SPAN DETECTION ISSUES}
\resizebox{\linewidth}{!}{%
\begin{tabular}{l|lll}
\hline
Disclosure Detection Span Issues                    & Accepted Spans (\texttt{495, 58\%}) & Rejected Spans (\texttt{356, 41.8\%}) & \begin{tabular}[l]{@{}l@{}}Total Occurrences (\texttt{868})\end{tabular} \\ \hline
Didn’t Make Sense               & \texttt{1 (0.2\%)}              & \texttt{115 (30.8\%)}           & \texttt{116 (13.4\%)}                                                                                         \\
Missing Context                 & \texttt{5 (1\%)}                & \texttt{28 (7.5\%)}             & \texttt{33 (3.8\%)}                                                                                           \\
Misaligned with User Preference            & \texttt{8 (1.6\%)}              & \texttt{152 (40.8\%)}           & \texttt{160 (18.4\%)}                                                                                         \\
** Incorrect Category Tag       & \texttt{33 (6.7\%)}             & \texttt{46 (12.3\%)}            & \texttt{79 (9.1\%)}                                                                                           \\
** Needs Multiple Category Tags & \texttt{2 (0.4\%)}              & \texttt{0}                      & \texttt{2 (0.2\%)}                                                                                            \\
Other (under/over-detected)  & \texttt{19 (3.8\%)}             & \texttt{9 (2.5\%)}              & \texttt{28 (3.3\%)}                                                                                           \\ \hline
\end{tabular}
}
\caption{This table summarizes the frequency of issues with disclosure detection spans that were reported by participants in our interviews. The disclosure detection spans are separated by whether they were ultimately accepted (495, 58\%) or rejected (356, 41.8\%) by participants as containing self-disclosures. The remaining 486 (57.1\%) of tags that contained no mistakes are not listed in this table. The most common issues among rejected tags were that they misaligned with users' understanding of sensitive information (40.8\%), and/or were identified as being nonsensical (30.8\%). While incorrect categorization was also an issue (affecting 9.1\% of all disclosure spans), it did not always lead to the rejection of a disclosure span (6.7\% were accepted). Note: Because there were overlaps in issues, the numbers for individual issues are greater than the number of tags (851). The total rates of accepted versus rejected highlights are adjusted based on this, while the total percentages of each mistake are based on a total number of occurrences (868).
\\***These issues are only relevant to the multi-class version of the model since the binary version did not contain disclosure categories.}
\end{table}

\subsection{RQ3: Where AI-powered Disclosure Detection Tools Can Fall Short And Be Improved}

We next analyzed our data to identify where participants felt our models fell short in order to identify opportunities for improvement. Our results surfaced many ways that automated disclosure detection tools --- like the models we developed in this work --- could be adjusted to better accommodate users' needs.

\paragraph{Misalignment of risk conception} One of the most common reasons users rejected model outputs was
that participants disagreed that a disclosure detected by the model was actually risky (see Table 3), and didn't mind disclosing that information resulting in a misalignment with user preference. This preference misalignment was the most common reason for rejection among nine of the seventeen disclosure categories detected by our model: i.e., ``relationship status'', ``pet'', ``finance'', ``age'', ``occupation'', ``education'', ``appearance'', ``husband/BF'', and ``wife/GF'' (see Table C in Appendix for the full summary of acceptance and rejection rates by category). 

For some detected disclosures, participants believed that the flagged disclosure could not actually be used to narrow down their identity to the extent that they could be re-identified: ``Okay, the next [disclosure span], ‘I've been happily monogamously remarried. Now just had our 15-year anniversary.’ I don't think that's identifiable at all. I think you know, lots of people have been married for 15 years. And nobody could really identify me from that specific information.'' (P7) This quote reaffirms the need for more granular explanations --- absent these explanations, users may fill the void with their own conception of attack possibilities which may not be fully representative of all possibilities. In this case, for example, while there may be many couples who have been married for 15 years (the absolute risk remains small), the number of people who were married exactly 15 years ago is much smaller than the number of, e.g., Reddit users as a whole (the relative risk is high). Combined with other identifying disclosures, the risk of re-identification may be higher than expected.

Yet, it is also important to not be overly prescriptive and to allow participants to exercise their own judgment. Even if there is an objective increase in risk and it is communicated well, sometimes participants viewed said risk as a necessary cost to communicate intent. As P4 states: ``I'm comfortable with what I've shared in the past and what I intend on sharing in the future...it adds value to my post, but in a way that doesn't personally identify me by any past nor any future posts that I intend to make.'' Any tool that aims to help users account for disclosure risks should allow users to make informed decisions --- not try to prevent users from taking any risks.

\paragraph{Missing disclosure categories} Participants also pointed out several kinds of disclosures that the model overlooked. Disclosures about \textit{illegal or abusive activities} --- such as harassment and substance abuse --- was one such category. In cases of harassment, for example, participants were worried that being too specific could open them up to re-identification by a perpetrator they were discussing. Referring to a post in which they were describing a situation with an estranged family member, P6 stated: ``I've been trying to keep my distance from him as much as possible and this seems like an invitation for him to go on a big rant defending himself, so I wouldn't want him to find it either...He would probably just try everything he could to contact me...and I already have him blocked on everything personal so he'd probably just message me a lot on this Reddit account and I might be tempted to just delete the account, or prevent him from contacting me forever.''

Participants also mentioned the importance of detecting disclosures related to \textit{addiction and substance use}. When discussing others' addictions --- e.g., in the case of a user seeking advice on how to help their spouse --- a primary concern was that the subject of the post might identify themselves if descriptions were too specific. On the other hand, when describing one's own addictions or substance-use habits, participants were more concerned that personal connections who knew of their account(s) might see this information attached to their profile via their posting history.

\textit{Ancestry and ethnicity} also came up as a category for disclosure detection models to identify. For example, one participant made a detailed post asking about a small town where her parent was born. They mentioned wanting this detected as a disclosure risk to protect themselves or others they discuss in their posts from further unwanted or unanticipated connections with past relationships: ``So somebody who had grown up there or had family who grew up there might know my family and be interested in more identifying information...it might be someone my dad doesn't like, and they might try to track him down and bother him further.'' (P19) They also mentioned concerns over other internet users reacting negatively when discussions around ethnicity arise: ``people do tend to get upset when ethnicity comes up. Discussions get heated...'' 

\textit{Temporal and situational details about specific events} (e.g. graduations, birthday parties), dates and times, instances of conflict (e.g., arguments about specific events), as well as direct quotes (from text messages, emails, or conversations on other platforms) were all also considered to be important categories of disclosures for the model to identify. These details can add a level of specificity to posts that made participants believe that any individuals involved in those events would easily be able to re-identify the post's author using those details.

In sum, these missing categories, and participants' rationales for their inclusion, underscore the need for disclosure detection models to account for participants' lived threat models --- some disclosures may be more risky for interpersonal threats (e.g., friends and relations learning personal information that the participant would rather keep private), others for third-party threats (e.g., strangers using disclosures to doxx an individual), and still others for institutional threats (e.g., law enforcement inferring illicit or illegal activities).

\paragraph{Accounting for disclosure norms and posting context} The third most common reason participants rejected model outputs ---  accounting for about 7.5\% of all rejections (see Table 3) --- was that the model failed to consider posting context. Several participants discussed situations where the detected disclosures were rendered irrelevant in light of the norms within the subreddit they were making the post: a common example was participants seeing location disclosures flagged when posting in a subreddit specifically about that location (e.g. ``my car was stolen from Tacoma, WA'' in r/Tacoma). Another example was detecting a disclosure about a poster's medical condition in a subreddit explicitly about that condition (e.g., sharing that one has Cushing's syndrome in r/Cushings). In other circumstances, certain personal disclosures were required by the subreddit's community guidelines, as was the case for P1 who had to disclose the name and age of her pet when posting to a subreddit about mourning family pets.  

\paragraph{Distinguishing between factual and non-factual disclosures} Our results also surfaced the importance of distinguishing between disclosures that were factual (e.g., about real people and situations) versus those that were not (e.g., when users intentionally perturbed identifying information, discussed hypotheticals, or discussed fictional characters). P10 shared: ``I mentioned...'suppose I had an office visit with a doctor'...it was clear that it wasn't even a real office visit, it was a hypothetical office visit to set the stage for the question to follow.'' 
This misunderstanding extended to other situations as well, such as when participants posted about games they were playing, with discussions of fictional characters being flagged with self-disclosure risks (e.g., the ``occupation'' of this character). Participants also described lying about details such as gender and age in their posts as a tactic for privacy risk management; in these cases, the model failed to account for these strategies and detected disclosure risks that participants already mitigated. P10 later reflected that this failure to distinguish between factual and non-factual disclosures was the model's most significant drawback: ``The biggest downside for these tools is that they're dumb. No awareness of context, no awareness of what I was actually trying to do in this post.'' 

\paragraph{Classification accuracy} Model accuracy was an overarching concern that often led to users' rejecting detected disclosures --- disclosure spans that did not make sense and erroneous category labels were the second most common reason users rejected model outputs (30.8\%) (see Table 3). Many erroneous disclosure spans, for example, consisted of one-off words such as ``the'', ``I'', or ``him''. Other errors included disclosures being mislabeled with the wrong class. We note, however, that among the 79 disclosures that were mislabeled with the wrong class, participants accepted 42\% of them (see Table 3, ``Incorrect Category Tag'' row). It is possible that even when the detected disclosure was mislabeled, it may have still brought users' attention to potentially risky disclosures that participants felt important to reconsider.

\paragraph{Calibrating detection frequency} Finally, over-detecting and under-detecting was also an issue for a minority (4/21) of participants, accounting for about 3.3\% (28/851) of all model outputs (see Table 3). In the over-detecting case, these participants expressed frustration that the model surfaced text that was \textit{irrelevant} to what participants identified as the disclosure, suggesting the disclosure spans be cut down the keywords that needed to be altered to improve the preservation of their anonymity (e.g. ``with her daughter in the back seat'' should be cut down to ``her daughter''). For under-detected passages, this occurred when participants considered them in the context of the sentence rather than in isolation. However, when discussing their overall outlook on the model, these issues did not always result in detected disclosures being rejected by participants --- only about 2.5\% (9/356) of detected disclosures that were considered redundant were ultimately rejected by participants (see Table 3).
\section{Discussion}
Our findings highlight both the promise and peril of using AI-assisted self-disclosure detection tools to help users make informed decisions about what to disclose when sharing personal information online. Whereas prior work in this space has focused on the \textit{technical} challenge of developing effective self-disclosure classification models, we show the problem is inherently socio-technical and emblematic of Ackerman's social-technical gap \cite{ackerman2000intellectual}. An effective solution cannot stop at achieving a high F$_1$ score; it will require significant consideration of, e.g., users' lived threat models, posting context, risk mitigation strategies, and understanding of risk. To that end, we culminate with a discussion on how such tools can be designed in a manner that is not only effective at detecting risky disclosures but useful in helping users make informed decisions.

To summarize our findings, the majority of participants responded positively to our model --- particularly the categorical disclosure variant --- describing its utility in facilitating self-reflection and in catching user's mistakes. Users also anticipated it being helpful in making decisions about content they might be uncertain about sharing. We also found that users appreciated the presence of coarse explanations provided by the categorical disclosure version of our model: since users' conceptions of risk may have differed from what the model was trained to identify, they were more likely to accept model outputs when given the ``category'' of the detected disclosure versus just the fact that a given span of text had some kind of disclosure. Finally, our findings also surface many weaknesses of existing models that highlight directions for future modeling and design work.

For future modeling work, our findings identify new categories of disclosure users desired the model to detect, the need to distinguish between factual and non-factual disclosures, to account for a user's history of disclosures, and the need to help users de-risk disclosures while preserving semantic meaning. For future design work, our findings suggest a need to design effective explanations, account for posting context and disclosure norms, and prioritize disclosures that are particularly important for users to heed so as to avoid habituation effects from over-warning users. Below, we center and embellish upon a few promising design implications for AI tools that aim to help users navigate self-disclosure risks online.

\subsection{Practical Implications for Deploying Self-Disclosure Detection Tools}
Our study raises important considerations for any practitioners or researchers hoping to apply the use of AI-assisted self-disclosure detection to existing systems, and showcases its potential utility across a multitude of contexts. Beyond being helpful as a standalone tool users could use across social media sites, we could also envision this being integrated into the design of platforms themselves, presented to users at the time of drafting a post. For example, such a tool could be smoothly integrated into existing features Reddit has released for making sure users adhere to community guidelines\footnote{Reddit announced the development of a post-guidance feature as part of their efforts to support moderators in sharing and enforcing ``community guidelines.'' This was noted in a post in r\/modnews, an official community dedicated to announcements from Reddit, Inc. pertaining to moderation: \url{https://www.reddit.com/r/modnews/comments/1cnacle/new_tools_to_help_mods_educate_and_inform/}}, though some work would be needed to ensure that the model suggestions complemented community guidelines as opposed to conflicting with them (e.g. including or excluding specific information from their posts).  AI-assisted self-disclosure detection can be especially useful to individuals seeking support in scenarios where discretion is crucial: for example, on platforms where domestic violence survivors are seeking support, where employees discuss how to navigate work-related issues (e.g. Blind), or even in contexts where whistle-blowers are reporting violations by companies or institutions. In the following paragraphs, we explore important considerations for any future NLP-based self-disclosure detection tools intended for use by end-users.

\paragraph{Surfacing only high-confidence outputs}
Our model made mistakes. While some participants found value in these mistakes in that the detected disclosures still drew their attention to other potentially risky disclosures in their posts,
too many mistakes reduced users' confidence in the effectiveness of the model overall.
Indeed, 29\% of participants either described not wanting to use the model, or being hesitant to use the model until performance improved.
Some participants also found that the model surfaced \textit{too many} potentially risky disclosures.
Prior work in usable privacy has shown that warnings can have a ``crying wolf'' effect --- i.e., habituation \cite{rankin2009habituation} --- where users react less strongly to warnings with each subsequent exposure.
Taken altogether, it would seem prudent for future work to explore better ways of handling model inaccuracies for disclosure detection tools. 

We envision two such approaches worthy of future investigation. 
First, researchers may want to explore other ways of presenting detected disclosure spans that lean further into the concept of self-reflection: i.e., users may be more willing to tolerate inaccuracies if detected disclosure spans are presented as suggestions for reflection rather than as definitive problems.
Second, given the strong negative impact of mistaken outputs future self-disclosure detection tools may only want to surface high-confidence decisions, so as not to break trust with users.
Future work can explore the appropriate confidence threshold, which may be variable by user, by disclosure category, and by threat model (e.g., users may be more willing to accept low confidence outputs when they feel in more immediate danger by a stalker or abusive ex-partner).

\paragraph{Concretizing Risk with Fine-Grained Explanations}
Participants expressed a preference for the categorical disclosure model outputs over the binary disclosure model outputs because the category labels gave them a sense of \textit{why} the model flagged a span as a potentially risky disclosure. The category labels, thus, served as a ``coarse'' explanation.
Future iterations of the model should explore more fine-grained explanations that help users understand and contextualize risk.
These fine-grained explanations may include, for example, estimates of k-anonymity loss, worst-case scenarios for how these disclosures could be used against users, or even a threat severity ranking for a given disclosure (each of which was requested in future iterations of the model by participants).
All of these explanations might help concretize the otherwise abstract risks of online self-disclosure.

\paragraph{Generative Models to Help Users De-Risk Disclosures}
Another idea worth exploring in future work is to provide users with suggestions on how to rephrase detected disclosures in a manner that preserves semantic meaning but de-risks the disclosure.
A naive approach would be to fine-tune or prompt-engineer a transformer-based model (e.g., an API-based model like ChatGPT \cite{gilardi2023chatgpt} or an open-source model like LLaMa-2 \cite{touvron2023llama}) to take an input span of text --- and the surrounding post context --- and generate suggested rephrasings of the text span that reduce disclosure risk. Future work should explore the relative costs and benefits of this naive approach versus a more involved approach in which the fine-tuned model more formally optimizes for an operationalized definition of privacy (e.g., k-anonymity loss) when generating suggested rephrasings.

Several participants also expressed a desire for suggested rephrasings to be personalized to their posting history. For example, some participants wanted the model to ``learn'' what they do and don't tend to disclose, providing suggestions based on these tendencies. Others described a desire for the suggestions and risks to be presented in the context of past disclosures that users made on the platform (e.g., ``Because you disclosed your age in a past post, sharing your location and occupation in this one will decrease your anonymity by x\%''). Further exploration of these features could also contribute to reduced cognitive load, and minimize aforementioned habituation effects raised by prior literature which are more common for generic and non-personalized warnings \cite{zimmermann2021nudge}. 

\paragraph{Accounting for posting context}
A core insight from our study is that it is critically important for disclosure detection models to account for posting context and the disclosure norms therein. The third most common reason for users' rejection of a detected disclosure was that the model failed to account for the context in which a detected disclosure span was being posted.

One way the model failed to account for context was by neglecting the subreddit in which a disclosure was being made.
Disclosing one's location in a subreddit about that location was one common example --- there is little point in hiding that one is in or from Texas when posting in a subreddit about Texas.
A second way the model failed to account for context was by neglecting disclosure norms.
In some subreddits, norms and guidelines necessitate certain disclosures (e.g., age and gender).
Disclosing this information is still risky, but participants mentioned that the model should at least account for this requirement and suggest alternative mitigation strategies (e.g., perturbing one's age within plausible bounds).

\paragraph{Responding to natural language instructions and feedback.}
Participants brought up a number of idiosyncratic weaknesses in the model outputs. For example, participants mentioned that the model failed to account for user's own privacy mitigation strategies (e.g., perturbing personal information in their posts), had difficulty distinguishing between hypothetical and real situations, missed some categories of disclosure that participants felt were important, and over-highlighted some disclosures that participants felt were not risky for their particular situation.
Addressing each of these weaknesses systematically at the model level would be difficult, but future work might explore the creation of instructable, transformer-based language models that can alter their behavior based on user-specified feedback and instructions (e.g., ``please don't surface disclosures related to age'', or ``this is a hypothetical situation that is not related to my personal information'').
Allowing for this sort of feedback may help reduce user frustration with model inaccuracies, especially if future models provide fine-grained explanations. Prior work has found that user frustration can increase when explanations are provided without giving users the ability to correct the model in cases of disagreement \cite{smith2020no}.

\section{Limitations}
\subsection{Model \& Study Design}
The disclosure spans detected by the binary and categorical disclosure variants of the model we used as a technology probe did not surface the exact same disclosure spans, reducing our ability to make direct comparisons between the two. We also did not counterbalance the appearance of the models --- users were always presented with the binary disclosure version of the model first. Thus reactions to the models may be subject to order effects. As such, we avoided making any direct statistical comparisons between the two models, instead opting to focus on \textit{why} users preferred one model's outputs over the other from their interview responses. Future work comparing the two versions of the model --- or alternatives with even finer-grained explanations --- might consider a between-subjects study design to eliminate concerns over priming and order effects. Due to the length of the interviews (1.5 hours) and the sheer number of disclosure spans users were being asked to assess, it was challenging to get the in-depth rationale for users' decisions behind their four ranks of each disclosure span while attempting to avoid over-fatigue on behalf of the participants. We also note that though we did probe users on their motivations for making posts, we did not observe any trends affecting users' assessments of model outcomes – though this could be due to the complexity of balancing the utility and needs for sharing with the risks. Future work could contribute a dedicated exploration of users' rationale for editing versus not editing disclosures that they agreed contained disclosure risks, as well as how their posting motivation might influence their perception of the model outcomes.

Relatedly, it is also important to note that users have differing threat models \cite{gallagher2017new, wu2018tree, redmiles2016think, zeng2017end} that may not align with the outputs of the model, and an important challenge exists in exploring how to best tailor the outputs of the model to meet users needs. That said, we believe this tool can still be particularly helpful for at-risk users (e.g. those seeking support in leaving abusive situations, reporting violations by companies, etc.)  

\subsection{Recruiting {\&} Sample Representation}
The gender demographics of our exploratory user study are significantly skewed toward people who identify as women. This gender distribution is not representative of the general population, nor is it representative of Reddit user gender statistics \cite{Liedke_Wang_2023}. Prior literature has well-established gender differences in privacy perceptions and behaviors \cite{hoy2010gender}, and as such our results may not be widely generalizable.
Our participant demographics are also skewed white, with a majority of participants below the age of 50, though these skews are more representative of Reddit demographics \cite{Liedke_Wang_2023}.
Furthermore, our sample was conducted solely with participants residing in the United States.
Though we attempted to recruit participants from the Reddit platform, we ultimately were only able to recruit participants from Prolific. Because we had used a newly created lab account to make recruitment posts, our posts were removed as our account ``did not have enough karma''. ``Karma'' is a form of social credit that is built on Reddit through commenting on others' posts, receiving upvotes, and awards; it is used to maintain the contribution quality of content posted to the site. While all interviewees were verified Reddit users, it's possible that we introduced additional biases into our sample as crowd-sourced participants are accustomed to participating and volunteering in research, and on the whole tend to be more tech-savvy than the general population \cite{redmiles2019well, peer2017beyond}.
Moreover, it's possible that those who responded to our recruitment call were more comfortable with sharing content that included personal disclosures online than others.
Finally, our work is highly specific to the Reddit context and, as such, our findings may not generalize to other pseudonymous or partially pseudonymous online fora (e.g., X, Mastodon).
\section{Conclusions}
When posting personal information in pseudonymous online fora, like Reddit, it is difficult for users to balance the tangible benefits of self-disclosure (e.g., providing context when seeking support) with its comparatively abstract risks (e.g., de-anonymization by institutions or third parties).
Prior work has explored the use of NLP-based self-disclosure identification tools to help users make more informed decisions by surfacing explicit disclosure risks in the content they plan to share online \cite{canfora2018nlp, morris2022unsupervised, guarino2022automatic}.
Building on this line of work, we contribute the first comprehensive evaluation of these tools with the users they aim to protect.
Through an in-depth interview study with {\N} Reddit users, using a state-of-the-art self-disclosure detection model as a technology probe \cite{dou2023reducing}, we explore if and how these models might be useful and usable beyond their F$_1$ scores.
Our findings suggest that many users find value in NLP-powered self-disclosure detection tools, particularly in their ability to facilitate self-reflection.
We also show that these models are especially effective when they provide users with context to interpret \textit{why} detected disclosures might be risky.
However, we also identified many shortcomings in these models that must be addressed to make them more useful and usable: for example, accounting for posting context, disclosure norms, and users' lived threat models.
Though modern advances in AI pose many thorny privacy risks for users \cite{lee2023deepfakes,das2023privacy}, our work shows one way these advances may be utilized to help users protect their privacy online.
\begin{acks}
This work was generously funded, in part, by the National Science Foundation under grants CNS-2316287, IIS-2144493, and IIS-2052498. We would also like to thank our anonymous reviewers for their helpful comments in revising this paper.
\end{acks}

\bibliographystyle{ACM-Reference-Format}
\bibliography{Paper-Refs}

\clearpage
\appendix

\appendix

\section{Appendices}
\subsection{Table A: Demographics}

\begin{table}[hbt!]
\caption*{DEMOGRAPHIC SUMMARY STRATIFIED BY GENDER}
\resizebox{0.9\linewidth}{!}{%
\begin{tabular}{@{}lllllll@{}}
\toprule
                                     &                           & Female  & Male       & Non-binary   & Total (N=21) \\ \midrule
\multirow{5}{*}{Age}                 & 18-29                     & \texttt{3 (25.0)}    & \texttt{3 (42.9)}    & \texttt{2 (100.0)}   & \texttt{8}   \\
                                     & 30-49                     & \texttt{6 (50.0)}    & \texttt{2 (28.6)}    & \texttt{0}           & \texttt{8}   \\
                                     & 50-64                     & \texttt{2 (16.7)}    & \texttt{1 (14.3)}    & \texttt{0}           & \texttt{3}   \\
                                     & 65 years and older        & \texttt{1 (8.3)}     & \texttt{1 (14.3)}    & \texttt{0}           & \texttt{2}   \\ \midrule
\multirow{4}{*}{Education}           & Graduate degree           & \texttt{1 (8.3)}     & \texttt{3 (42.9)}    & \texttt{1 (50.0)}    & \texttt{9}   \\
                                     & Bachelor's degree         & \texttt{3 (25.0)}    & \texttt{3 (42.9)}    & \texttt{0}           & \texttt{6}  \\
                                     & Some college              & \texttt{7 (58.3)}    & \texttt{1 (14.3)}    & \texttt{0}           & \texttt{8}  \\
                                     & High school degree        & \texttt{1 (8.3)}     & \texttt{0}           & \texttt{0}           & \texttt{1}  \\ \midrule
\multirow{6}{*}{Employment Status}   & Full-time                 & \texttt{1 (8.3)}     & \texttt{3 (42.9)}    & \texttt{0}           & \texttt{4}  \\
                                     & Part-time                 & \texttt{5 (41.7)}    & \texttt{1 (14.3)}    & \texttt{1 (50.0)}    & \texttt{7}  \\
                                     & Unemployed \& looking     & \texttt{2 (16.7)}    & \texttt{1 (14.3)}    & \texttt{1 (50.0)}    & \texttt{4}   \\
                                     & Unemployed \& not looking & \texttt{3 (25.0)}    & \texttt{2 (28.6)}    & \texttt{0}           & \texttt{5}   \\
                                     & Self-employed             & \texttt{1 (8.3)}     & \texttt{0}           & \texttt{0}           & \texttt{5}   \\ \midrule
\multirow{5}{*}{Income}              & \$100k+                  & \texttt{0}           & \texttt{2 (28.6)}    & \texttt{0}           & \texttt{2}   \\
                                     & \$75k-99k                 & \texttt{1 (8.3)}     & \texttt{2 (28.6)}     & \texttt{0}          & \texttt{3}   \\
                                     & \$50k-74k                & \texttt{2 (16.7)}    & \texttt{2 (28.6)}    & \texttt{1 (50.0)}    & \texttt{5}   \\
                                     & \$25k-49k                 & \texttt{3 (25.0)}    & \texttt{1(14.3)}     & \texttt{0}           & \texttt{4}   \\
                                     & \textless\$25k            & \texttt{6 (50.0)}    & \texttt{0}           & \texttt{1 (50.0)}    & \texttt{7}   \\ \midrule
\multirow{3}{*}{Ethnicity}           & Black/African             & \texttt{2(16.7)}     & \texttt{1(14.3)}     & \texttt{0}           & \texttt{3}   \\
                                     & Caucasian                 & \texttt{10 (83.3)}   & \texttt{4 (57.1)}    & \texttt{2 (100.0)}   & \texttt{16}  \\
                                     & South Asian               & \texttt{0}           & \texttt{2 (28.6)}    & \texttt{0}           & \texttt{2}   \\ \midrule
\multirow{4}{*}{Location}            & Large city                & \texttt{7 (58.3)}    & \texttt{2(28.6)}     & \texttt{1 (50.0)}    & \texttt{10}  \\
                                     & Suburb near large city    & \texttt{3 (25.0)}    & \texttt{4 (57.1)}    & \texttt{1 (50.0)}    & \texttt{8}   \\
                                     & Small city or town        & \texttt{1 (8.3)}     & \texttt{1 (14.3)}    & \texttt{0}           & \texttt{2}   \\
                                     & Rural area                & \texttt{1 (8.3)}     & \texttt{0}           & \texttt{0}           & \texttt{1}   \\  \bottomrule
\end{tabular}%
}
\end{table}

\clearpage

\subsection{Table B: User Alteration Decisions}

\begin{table}[hbt!]
\caption*{SUMMARY OF USER REACTIONS TO ACCEPTED VS. REJECTED DISCLOSURE SPANS}
\centering
\resizebox{0.7\linewidth}{!}{%
\arrayrulecolor{black}
\begin{tabular}{l!{\vrule}l!{\vrule}l!{\vrule}l!} 
\hline
                  & Accepted Spans (495) & Rejected Spans (356) & Total (851)  \\ 
\hline


**Altered\textit{~} & \texttt{23\% (114)}              & \texttt{3.7\% (13)}                 &                   \\ 

Undecided         & \texttt{5\% (23)}                &\texttt{2\% (7)}                    & \texttt{3\% (30)}          \\ 

Don’t Alter       & \texttt{72\% (357) }             & \texttt{94.1\% (335) }              &\texttt{81\% (692) }        \\
\hline
\end{tabular}
}
\caption{The rate of different reactions participants had to the disclosure detection spans, compared between disclosure spans that participants agreed contained self-disclosure (accepted spans), and spans participants felt did not contain self-disclosure (rejected spans). \\**The altered rate is a sum of changed spans and removed disclosure spans.}

\end{table}

\subsection{Table C: Categorical Disclosure Model Disclosure Span Acceptance/Rejection by Category}

\begin{table}[hbt!]
\caption*{SUMMARY OF USER REACTIONS \& DETECTION ISSUES ACROSS DIFFERENT CATEGORIES OF MULTI-CLASS MODEL}
\centering
\resizebox{\linewidth}{!}{%
\begin{tabular}{l|lll|llllll}
\hline
\begin{tabular}[l]{@{}l@{}}Category \\ (By most populous)\end{tabular} & \begin{tabular}[l]{@{}l@{}}Rejection \\ Rate\end{tabular} & \begin{tabular}[l]{@{}l@{}}Acceptance \\ Rate\end{tabular} & Count  & \begin{tabular}[l]{@{}l@{}}Didn’t \\ Make Sense\end{tabular} & \begin{tabular}[l]{@{}l@{}}Missing \\ Context\end{tabular} & \begin{tabular}[l]{@{}l@{}}User Preference \\ Misalignment\end{tabular} & \begin{tabular}[l]{@{}l@{}}Incorrect \\ Tag\end{tabular} & \begin{tabular}[l]{@{}l@{}}Multiple \\ Tags\end{tabular} & \begin{tabular}[l]{@{}l@{}}Other \\ (under/over highlighted)\end{tabular} \\ \hline
Family                                                                 & \texttt{35\% }                                                     & \texttt{65\%}                                                       & \texttt{158}    & \texttt{28 (18\%)}                                                    & \texttt{8 (5\%)}                                                    & \texttt{14 (9\%)}                                                       & \texttt{35 (22\%)}                                                & \texttt{1 (1\%)}                                                  & \texttt{7 (4\%)}                                                                   \\
Husband/BF                                                             & \texttt{43\%}                                                      & \texttt{57\%}                                                       & \texttt{81}     & \texttt{7 (9\%)}                                                      & \texttt{2 (2\%)}                                                    & \texttt{21 (26\%)}                                                      & \texttt{13 (16\%)}                                                & \texttt{0}                                                        & \texttt{2 (2\%)}                                                                   \\

Health & \texttt{31\%} & \texttt{69\%} & \texttt{58} & \texttt{12 (21\%)} & \texttt{3 (5\%)} & \texttt{3 (5\%)} & \texttt{6 (5\%)} & \texttt{0} & \texttt{0}
                                                                       \\
Mental Health & \texttt{27.5\%} & \texttt{72.5\%} & \texttt{40} & \texttt{8 (20\%)} & \texttt{0} & \texttt{2 (5\%)} & \texttt{6 (15\%)} & \texttt{0} & \texttt{2 (5\%)} \\

Location & \texttt{24\%} & \texttt{76\%} & \texttt{34} & \texttt{1 (3\%)} & \texttt{0} & \texttt{5 (15\%)} & \texttt{1 (29\%)} & \texttt{0} & \texttt{0} \\

Relationship Status & \texttt{38\%} & \texttt{62\%} & \texttt{34} & \texttt{3 (9\%)} & \texttt{0} & \texttt{4 (12\%)} & \texttt{3 (9\%)} & \texttt{0} & \texttt{1 (2\%)} \\

Pet                                                                    & \texttt{28\%}                                                      & \texttt{72\%}                                                       & \texttt{32}     & \texttt{0}                                                            & \texttt{1 (3\%)}                                                    & \texttt{5 (16\%)}                                                       & \texttt{5 (6\%)}                                                  & \texttt{0}                                                        & \texttt{3 (9\%)}                                                                   \\
Finance                                                                & \texttt{46\%}                                                      & \texttt{54\%}                                                       & \texttt{24}     & \texttt{4 (17\%)}                                                     & \texttt{1 (4\%)}                                                    & \texttt{4 (17\%)}                                                       & \texttt{1 (4\%)}                                                  & \texttt{0}                                                        & \texttt{0}                                                                         \\
Age                                                                    & \texttt{38\%}                                                      & \texttt{62\%}                                                       & \texttt{21}     & \texttt{2 (9\%)}                                                      & \texttt{2 (10\%)}                                                   & \texttt{3 (14\%)}                                                       & \texttt{1 (5\%)}                                                  & \texttt{0}                                                        & \texttt{0}                                                                         \\

Occupation                                                             & \texttt{45\%}                                                     & \texttt{55\%}                                                       & \texttt{20}     & \texttt{1 (5\%)}                                                      & \texttt{1 (5\%)}                                                    & \texttt{5 (25\%)}                                                       & \texttt{5 (25\%)}                                                  & \texttt{0}                                                        & \texttt{2 (10\%)}                                                                  \\
Age/Gender                                                             & \texttt{12.5\%}                                                    & \texttt{87.5\%}                                                     & \texttt{16}     & \texttt{0}                                                            & \texttt{0}                                                          & \texttt{3 (19\%)}                                                       & \texttt{0}                                                        & \texttt{0}                                                        & \texttt{0}                                                                         \\
Gender                                                                 & \texttt{30\%}                                                      & \texttt{70\%}                                                       & \texttt{10}     & \texttt{2 (20\%)}                                                     & \texttt{2 (20\%)}                                                   & \texttt{0}                                                              & \texttt{3 (30\%)}                                                 & \texttt{0}                                                        & \texttt{0}                                                                                                                                               \\
Sexual Orientation                                                     & \texttt{50\%}                                                      & \texttt{50\%}                                                       & \texttt{10}     & \texttt{3 (30\%)}                                                     & \texttt{0}                                                          & \texttt{1 (10\%)}                                                       & \texttt{4 (40\%)}                                                 & \texttt{0}                                                        & \texttt{0}                                                                         \\
Education                                                              & \texttt{25\%}                                                      & \texttt{75\%}                                                       & \texttt{8}      & \texttt{0}                                                            & \texttt{0}                                                          & \texttt{2 (25\%)}                                                       & \texttt{0}                                                        & \texttt{1 (13\%)}                                                 & \texttt{0}                                                                         \\
Appearance                                                             & \texttt{57\%}                                                      & \texttt{43\%}                                                       & \texttt{7}      & \texttt{0}                                                            & \texttt{0}                                                          & \texttt{1 (14\%)}                                                       & \texttt{1 (14\%)}                                                 & \texttt{0}                                                        & \texttt{0}                                                                         \\

Race/Nationality                                                       & \texttt{0\%}                                                       & \texttt{100\%}                                                      & \texttt{4}      & \texttt{0}                                                            & \texttt{0}                                                          & \texttt{0}                                                              & \texttt{1 (25\%)}                                                 & \texttt{0}                                                        & \texttt{0}                                                                         \\
Wife/GF                                                                & \texttt{100\%}                                                     & \texttt{0\%}                                                       & \texttt{4}      & \texttt{2 (50\%)}                                                     & \texttt{0}                                                          & \texttt{2 (50\%)}                                                       & \texttt{0}                                                        & \texttt{0}                                                        & \texttt{0}                                                                         \\
\hline
                                                                       &                                                           &                                                            & Total: & \texttt{73}                                                           & \texttt{20}                                                         & \texttt{75}                                                             & \texttt{69}                                                       & \texttt{2 }                                                       & \texttt{17}                                                                        \\ \hline
\end{tabular}
}
\caption{The rate of acceptance and rejection across different categories of disclosure detected in the multi-class model, alongside rates of model detection issues. \\Note: This table only displays statistics for the multi-class model. "Name" and "Contact Information" do not appear in this table since they did not appear in any post detections.}
\end{table}
\clearpage

\subsection{Figure 4: Likert Scale Wording for Disclosure Detection Rankings}
\begin{table*}[htp]
        \begin{tabular}{ll}
            \toprule
            How would you rate the \textbf{helpfulness} of this highlight in surfacing a self-disclosure risk?  \\
            \midrule
            (1) Not at all helpful \\
            (2) Somewhat helpful  \\
            (3) helpful \\
            (4) Fairly helpful  \\
            (5) Very helpful  \\
            \midrule
            How \textbf{important} was it to disclose that information highlighted in your post? \\
            \midrule
            (1) Not at all important \\
            (2) Somewhat important  \\
            (3) Important \\
            (4) Fairly important  \\
            (5) Very important  \\
            \midrule
            How would you rate the \textbf{sensitivity} of the content highlighted? \\
            \midrule
            (1) Not at all sensitive\\
            (2) Somewhat sensitive\\
            (3) Sensitive \\
            (4) Fairly sensitive\\
            (5) Very sensitive\\
            \midrule
            How \textbf{risky} do you feel it is to disclose the information highlighted in an  online post? \\
            \midrule
            (1) Not at all risky\\
            (2) Somewhat risky\\
            (3) Risky\\
            (4) Fairly risky\\
            (5) Very risky\\
            \midrule
            \bottomrule
            
        \end{tabular}
            \caption{"Helpfulness", "importance", "sensitivity", and "riskiness" all ranked on a 5-point Likert Scale."}
\end{table*}
\clearpage

\subsection{Figure 5: Results of the Mann-Whitney U Test Alongside Welch's T-test}
\begin{table}[htp]
\centering
\caption*{SUMMARY STATISTICS FOR ACCEPTED SELF-DISCLOSURE SPANS OUTPUT BY THE MODELS}
\resizebox{\linewidth}{!}{%

\begin{tabular}{lcc|cc|ll|ll} 
\hline\hline
\multicolumn{1}{c}{} & \multicolumn{2}{c|}{\begin{tabular}[c]{@{}c@{}}Accepted Altered Disclosure Spans\\ (114)\end{tabular}} & \multicolumn{2}{c|}{\begin{tabular}[c]{@{}c@{}}Accepted Non-Altered Disclosure Spans \\ (357)\end{tabular}} & \multicolumn{2}{c|}{\begin{tabular}[c]{@{}c@{}}Welch's T-Test Results \\ 
\end{tabular}} & \multicolumn{2}{c}{Mann-Whitney U Results}  \\ 

\cline{2-9}
                     & M    & SD                                                                                        
                     & M    & SD                                                                                          & p-value      & Cohen's \textit{d }        
                     & p-value \\ 
\hline
Helpfulness          & \texttt{4.14} & \texttt{1.24}                                                                                            & \texttt{3.15} & \texttt{1.44}                                                                                                 & \texttt{1.52702e-10***} & \texttt{0.24}       
                                        & \texttt{1.489305e-09***}                                   \\
Importance           & \texttt{2.58} & \texttt{1.53}                                                                                            & \texttt{3.74} & \texttt{1.25}                                                                                                 & \texttt{7.728284e-11***} & \texttt{0.45}              
                                        & \texttt{3.971501e-12***}                                   \\
Sensitivity          & \texttt{3.95} & \texttt{1.36}                                                                                            & \texttt{2.62} & \texttt{1.39}                                                                                                 & \texttt{1.884628e-15***} & \texttt{0.52 }                  
                                        & \texttt{1.85884e-14***}                                   \\
Riskiness            & \texttt{3.61} & \texttt{1.47}                                                                                            & \texttt{2.22} & \texttt{1.35}                                                                                                 & \texttt{8.741327e-15***} & \texttt{0.52}                   
                                        & \texttt{4.353914e-15***}                                    \\
\hline
\end{tabular}
}
\caption{Results of both the Mann-Whitney U test and Welch's T-test for disclosure spans accepted by participants to explain \emph{nuances in when participants altered and did not alter their posts} in response to seeing the model outputs. The outcomes of the two tests are in agreement, showing significant differences in the mean and median scores of disclosure spans between those accepted and altered compared to those accepted but not altered by participants across helpfulness, importance, sensitivity, and riskiness.
\\M = Mean, SD = Standard Deviation
\\Helpfulness, Importance, Sensitivity, and Riskiness all range from 1 (“Not at all...”) to 5 (“Very...”) 
\\Note: p-values for both the Welch's t-test and the Mann-Whitney U test have been adjusted for multiple hypothesis testing. 
\\ *\emph{p}<0.05, **\emph{p}<0.01, ***\emph{p}<0.001}
\end{table}

\end{document}